\documentclass[12pt]{spieman}  
\usepackage{amsmath,amsfonts,amssymb}
\usepackage{graphicx}
\usepackage{setspace}
\usepackage{tocloft}
\usepackage{booktabs}
\usepackage{multirow}
\usepackage{array}
\usepackage[numbers]{natbib}
\usepackage{breqn}
\usepackage{environ}
\usepackage{wrapfig}
\usepackage{makecell}
\usepackage{rotating}
\usepackage{lineno}
\usepackage{xcolor}
\NewEnviron{LargerTxt}{%
    \scalebox{1.2}{$\BODY$}
}

\title{Shedding Light on Large Space-Based Telescopes: Modeling Stray Light due to Primary Mirror Damage from Micrometeoroid Impacts}

\author[1,2*]{Megan T. Gialluca}
\author[3]{Jonathan W. Arenberg}
\author[4,5]{Chris Stark}
\author[3]{Blake Shepherd}
\author[1,2,6]{Victoria S. Meadows}
\author[5]{Aki Roberge}
\author[7]{Tyler D. Robinson}
\author[5]{Robert Podgurski}
\affil[1]{Department of Astronomy and Astrobiology Program, University of Washington, Box 351580, Seattle, Washington 98195, USA}
\affil[2]{NExSS Virtual Planetary Laboratory, Box 351580, University of Washington, Seattle, Washington 98195, USA}
\affil[3]{Northrop Grumman Systems Corporation, Redondo Beach, California, United States}
\affil[4]{Exoplanets and Stellar Astrophysics Lab, Code 667, NASA Goddard Space Flight Center, Greenbelt, MD 20771, USA}
\affil[5]{Astrophysics Division, NASA Goddard Space Flight Center, Greenbelt, MD 20771, USA}
\affil[6]{SETI Institute, 339 Bernardo Ave, Suite 200, Mountain View, California, 94043, USA}
\affil[7]{Lunar and Planetary Laboratory, University of Arizona, Tucson, AZ 85721, USA}

\newcommand{\revision}[1]{\textcolor{black}{#1}}

\cftpagenumbersoff{figure}
\cftpagenumbersoff{table} 
\begin{document} 
\maketitle


\begin{abstract}
A large space-based telescope aimed at detecting and characterizing the atmospheres of Earth-like planets orbiting Sun-like stars will require unprecedented contrast and stability. However, damage to the primary mirror due to micrometeoroid impacts will provide a stochastic, time-dependent source of stray light in the coronagraph's field of view that could significantly lengthen exposure times and reduce the expected science yield. To better quantify the impact of stray light and inform the Habitable Worlds Observatory mission design process, we present estimates of stray light in different micrometeoroid damage scenarios for a broad range of targets, and use that to find the expected decrease in science yield (i.e., the expected number of detected exoEarth candidates). We find that stray light due to micrometeoroid damage may significantly reduce yield, by 30\% -- 60\% in some cases, but significant uncertainties remain due to the unknown maximum expected impactor energy, and the relationship between impact energy and expected crater size. Micrometeoroid damage therefore needs further exploration, as it has the potential to reduce scientific yield, and in turn drive the development of mitigation strategies, selection of telescope designs, and selection of observing priorities in the future.
 
\end{abstract}

\keywords{stray light, telescopes, mirrors, exoplanets}

{\noindent \footnotesize\textbf{*}Megan T. Gialluca,  \linkable{gialluca@uw.edu} }

\vspace{10mm}
\noindent
Accepted for Publication in SPIE JATIS

\begin{spacing}{2}   

\section{Introduction}
\label{sect:intro}

The recent prioritization of the Habitable Worlds Observatory as NASA's next Astrophysics flagship mission will significantly advance the search for life on exoplanets.   
In 2021, the \textit{Pathways to Discovery} decadal survey recommended the Habitable Worlds Observatory (HWO) as the first mission to enter the Great Observatories Maturation Program (GOMAP) in the 2020s \citep{national2021decadal}. HWO has been charged with surveying at least 100 nearby Sun-like stars to detect potential exoEarth candidates (i.e., habitable zone terrestrial planets), and subsequently following up on the most exciting 25 planets to collect spectra for atmospheric characterization \citep{national2021decadal}. This mission will require unprecedented contrast and stability, as well as a large ($\geq$6 m diameter) primary mirror, and will use coronagraph technology for starlight suppression \citep{national2021decadal}. Furthermore, the mission will be focused on UV, optical, and near-infrared wavelengths, setting it apart from the infrared wavelength regime of JWST. 

In general, space telescopes have to account for \revision{numerous sources of stray light, which will reduce the sensitivity of scientific observations. These sources may include scatter from the mirrors, coatings, and chamfers in addition to contamination, and damage to the optics, such as micrometoroid impact craters on the primary mirror.}
\citep[e.g.,][]{Arenberg2019turningrocks,Rigby2022characterization,Rigby2023dark}. For a given mission, some of these sources will tend to dominate over others, due to the technology used for that mission and the wavelengths of interest; for example, the JWST--- operating in the near- to mid-infrared---is primarily limited by zodiacal emission \citep{Rigby2022characterization}. 

Though micrometeoroid impacts are known to affect space telescopes, very few studies have considered the contribution of stray light from micrometeoroid damage, and in general these studies have found it to be a negligible source of noise \citep{Lightsey2012jwstStrayLightwDamage,Khodnevych2020LISAMeteor,Sang2023straylightTaijiTelescope}. 
However, because HWO will routinely be attempting to characterize ultra-faint objects of 30th magnitude and fainter \citep{Stark2024yield}, exposure times will be heavily influenced by stray light, including that contributed by micrometeoroid damage to the primary mirror. The LUVOIR mission concept studied the use of a coronagraph to facilitate the detection of exoEarth candidates, which suggested this search should be performed with a visible wavelength coronagraph covering a bandpass of 500 nm -- 1000 nm to target the water vapor feature at 950 nm and an oxygen feature at 760 nm \citep{luvoir2019luvoir,Stark2024yield};
\revision{more recent work focused on HWO has also confirmed the suitability of the 500 and 1000 nm bandpasses \cite{stark2024optimized}.}
\revision{Visible wavelength observations also allow us to operate at a larger working angle (in units of $\lambda$/D) compared to a longer wavelength feature \citep{luvoir2019luvoir}.} However, shorter wavelengths make stray light a greater concern because Rayleigh scattering scales with $\lambda^{-4}$. Moreover, though coronagraphs are useful for revealing the ultra-faint planet beside a bright host star, coronagraphs cannot remove diffuse light scattered into the observation in the same way \citep{Balasubramanian2009coronagraphperform,Pfisterer2018scatterexoplanets,Breckinridge2018pinwheelpupil}. 

In this work, we demonstrate that stray light due to micrometeoroid damage has the potential to significantly affect the mission. \revision{We focus our efforts on the uncertainties in micrometeoroid flux and impact crater size and its effect on stray light. We will show that science mission performance (e.g., yield) is sensitive to this stray light and the current level of uncertainty results in considerable variability in science performance. To have an efficient and determinative design process the impact of damage on science performance must be much better understood that is the current state of the art.} 
To conduct this study, we develop a basic stray light model that employs analytical and numerical integration for increased computational efficiency over more sophisticated ray tracing methods. This approach allows for the exploration of a wide range of mission considerations such as telescope barrel design, target pool selection, and micrometeoroid impact scenarios. We then approach this problem by first quantifying the order of magnitude of stray light that may be expected from single micrometeoroid impacts to the primary mirror for a range of impact energies. We then directly demonstrate the effect of micrometeoroid damage on the science goals of HWO by computing the expected reduction in detected exoEarth candidate yield resulting from increased exposure times following various single-hit impact events. \revision{While this study is focused on the effects of single-hit impactors, we also note that the build-up of small, low-energy impacts over time may also contribute to stray light concerns for the mission. Although this build-up of small impact events is not in the scope of this work, the model developed here could be used to conduct a more detailed study on this subject in the future.} 

In the following section (\S\ref{sec:analyticApproach}), we introduce the analytic approach to our simple model of stray light from primary mirror defects, including defining and exploring bi-directional reflectance distribution (BRDF) models which describe the reflectance properties of various idealized mirror imperfections. In section \ref{sec:modelmethods}, we discuss the computational methods applied to model stray light from a resolved sky background onto a resolved mirror surface across, as well as the tools used to calculate exoEarth candidate yield \citep{Stark2014AYO,Stark2019yieldandrevisits,Stark2024yield}. Section \ref{sec:Results} presents the results of this work, emphasizing the relationship between micrometeor impact energy and exoEarth candidate yield. Finally we discuss the implications of our results for the goals of HWO in \S\ref{sec:discussion}.

\section{Analytical Methods} \label{sec:analyticApproach}


We will estimate levels of stray light due to scatter from  micrometeoroid-formed craters (particularly from hyper-velocity impacts) in the excavated and distorted material and coatings of a mirror. By approximating impact craters and ejected contamination as a dig in the surface quality grading, we can then apply the Peterson reflection model \citep{Peterson2012brdf} as an idealized functional form describing the attributes and scale of a dig caused by micrometeoroid collisions. From this form, the BRDF is derived and applied to estimate the scalar amount of light scattered at points in the image plane away from the ideal point spread function (or image) of that same mirror undamaged; that is, a point spread function composed of specular and diffuse reflections only.

We first consider the host star as the sole stray light source, and then expand the  stray light source to the spatially-resolved sky background. To describe scattering, we first consider perfect specular and diffuse (Lambertian) reflection (\S \ref{subsubsec:diffuseVersSpec}) and then employ the Peterson reflection model for digs (i.e., dents on a mirror) \citep{Peterson2012brdf} to represent micrometeoroid impact crater damage (\S \ref{subsubsec:PetersonModel}). \revision{Finally, we will generalize the equations to describe stray light from a multiple source, spatially resolved sky background onto a spatially resolved (distributed impact damage from a given event) telescope mirror (\S \ref{subsec:generalized}),  which is used in our computational model for stray light from a resolved sky background onto a  resolved mirror surface in the next section (\S \ref{sec:modelmethods}).}

\subsection{An Initial Estimate}
\label{subsec:Estimate}

\revision{First, we isolate the host star as the only stray light source of photon flux on the mirror. We assume the host star flux is reflected by damaged mirror (incoherent and uncontrollable) into the image plane, and the total stray light flux in the observation is found by integrating the BRDF across the size and location of the planet's PSF core.} This stray light is used to calculate the allowable area of damage to the mirror to not exceed a specified stray light requirement. 
The total signal in the stray light flux at the location of a targeted planet's own PSF is simply the integration of the BRDF over the equivalent solid angle of the planet's PSF core, at the angular location of the planet in the image plane.

To begin with a strict fiducial scenario, we assume the stray light scattered into the planet's PSF core must be less than the light from the planet being observed. We select this fiducial scenario because the exposure times required to observe a planet will increase after this point, but note that it is likely more strict than needed for the mission. This requirement can be represented with the following equation:
\begin{equation} \label{eq:StartFullSum}
    F_{\rm p} \ A_{PM} \ cos(\theta_{i}) \geq F_* \ \sum_k \left( A_{D,k} \ cos(\theta_i) \ \int_{\omega_r} \ f_{r,k}(\theta_{i}, \phi_{i}, \theta_{r}, \phi_{r}) \ cos(\theta_{r}) \ d\omega_r \right)\ .
\end{equation}
Where F$_{p}$ is the flux of the planet (W/m$^{2}$/$\mu$m), F$_{*}$ is the flux of the host star (W/m$^{2}$/$\mu$m), A$_{PM}$ is the collecting area of the primary mirror (m$^{2}$), 
A$_{D,k}$ is the defective area for defect point $k$ on the mirror (m$^{2}$), and $f_{r,k}(\theta_{i}, \phi_{i}, \theta_{r}, \phi_{r})$ is the bi-directional reflectance distribution function, or BRDF (1/steradians), for defect $k$ in the direction of reflection we are considering, which is defined by the angles of incidence and reflection ($\theta_i$ and $\theta_r$) and the azimuthal angles to the direction of incidence and reflection ($\phi_i$ and $\phi_r$). \revision{We assume all of the planetary flux reflected off the primary mirror lands in the PSF core.} When the BRDF $f_{r,k}(\theta_{i}, \phi_{i}, \theta_{r}, \phi_{r})$ is then integrated across a solid angle in some reflection direction ($\omega_r$), the resulting value is the fraction of F$_{*}$ incident on $A_{D,k}$ that is reflected into the solid angle, $\omega_r$. In this case, $\omega_r$ would represent the size of the core of the planet's PSF, which we will take to have a radius of $\lambda$/D. Primary mirror area is also multiplied by the cosine of the angle of incidence to account for the reduction of true area to projected area; however, for this case of only a planet and its host star providing light, the angles of incidence and reflectance are 0$^{o}$ and the cosine term becomes 1. 

For a more simplified approximation, we will treat the mirror as a point source surface and assume the sum over the types of defective area times the integral of the associated BRDF over all applicable scattering directions reduces to some total defective mirror area with some average BRDF:
\begin{equation} \label{eq:AvgDamage}
    F_p A_{PM} \geq F_* \  A_{D} \ \int_{\omega_r} \left( f_r(\theta_{i}, \phi_{i}, \theta_{r}, \phi_{r}) \ cos(\theta_r) \ d\omega_r \right) \ .
\end{equation}
We can then then rearrange Equation \ref{eq:AvgDamage} to solve for the fraction of allowed defective mirror area:
\begin{equation} \label{eq:nonsimplifiedBRDF}
    \frac{A_{D}}{A_{PM}} \leq \ \left[ \int_{\omega_r} \left( f_r(\theta_{i}, \phi_{i}, \theta_{r}, \phi_{r}) \ cos(\theta_r) \ d\omega_r \right) \right]^{-1} \ \frac{F_p}{F_*} \ .
\end{equation}
Thus, Equation \ref{eq:nonsimplifiedBRDF} gives the fraction of allowed defective mirror area (A$_{D}$/A$_{PM}$) for a particular assumed BRDF, assuming the host star is the only stray light source and our maximum stray light requirement is the received flux of the planetary companion of interest within the area of its PSF. While we begin with this requirement, we stress that the adopted stray light requirement may be too strict, especially given that exozodiacal dust may be the typical driver of exposure times \citep{Mennesson2024exozodi} which can be brighter than the planet within the PSF area. In our results section (\S \ref{sec:Results}) we will remove this requirement, imposing no maximum limit on stray light, and allow exposure times to increase with stray light to show the effect on science yield.

\subsubsection{Diffuse vs. Specular Reflection} \label{subsubsec:diffuseVersSpec}


As a first step in determining the relative importance of diffuse and specular components of stray light from a defective mirror area we explore the extremes in assuming that the defective mirror area is either a perfectly diffuse reflector, or a near-perfect specular reflector. 

If we first consider the defective area to be a diffuse reflector, we can adopt a Lambertian diffuse scattering BRDF with a diffusive reflectance of $\rho_d$ into Equation \ref{eq:nonsimplifiedBRDF}; consequently, the integral over $\omega_r$ reduces to multiplying by the solid angle of the core of the planet's PSF ($\Omega_{PSF}$):
\begin{equation} \label{eq:ImperfectArea}
    \frac{A_{D}}{A_{PM}} \leq \frac{1}{\Omega_{PSF}} \ \frac{\pi}{\rho_d} \ \frac{F_p}{F_*} \ .
\end{equation}

Equation \ref{eq:ImperfectArea} assumes that all flux from the planet incident on the primary mirror is perfectly reflected and focused to a point source on the image plane. In contrast, flux from the host star hitting the defective mirror area is diffusely scattered according to a Lambertian (cosine) distribution, spreading the light evenly across the image plane.
The ratio of Earth to Sun flux in the V band at full phase is given by:
\begin{equation} \label{eq:EarthSunVband}
    \frac{F_{\oplus}}{F_{\odot}} = A_{g}(\nu) \left( \frac{R_{\oplus}}{a} \right)^{2} = 0.2 \left( \frac{1 R_{\oplus}}{1 AU} \right)^{2} = 3.6 \revision{\times} 10^{-10} \ ,
\end{equation}
\revision{which may simply be taken as 1$\times$10$^{-10}$. Here, $A_g(\nu)$ is the geometric albedo, and $a$ is the semi-major axis of Earth.} The radius of the core of the planet's PSF in the V band ($\lambda_{eff}$ = 0.545 $\mu$m) for a telescope with a 6 m diameter would be $\sim$9$\times$10$^{-8}$ radians, \revision{which yields a solid angle of:
\begin{equation}
    \Omega_{PSF} = 2 \pi (1 - cos(\lambda/D)) \approx 3\times10^{-14} \ sr \ .
\end{equation}}
If we assume all light is reflected, and all reflection is diffuse, then $\rho_d$ = 1. If we plug these values into Equation \ref{eq:ImperfectArea}, the allowed defective mirror area \revision{would be $>$10$^{4}$ times} the area of the primary mirror. This allowed damage would only increase if $\rho_d$ was taken to be less than 1. \revision{While this result mathematically works out, the physical interpretation is confusing as a mirror cannot be damaged more than 100\%. This result arises from the assumption that the full mirror area always reflect planetary light properly, no matter how much damage has been sustained. Ordinarily this assumption is fine, as the damaged area is typically much less than the full mirror, but in this case, a more physically realistic result should take into account that the planetary light will not be properly reflected by damaged area.}

\revision{Under the more realistic treatment that accounts for the decreasing area available to reflect planetary light as damage is sustained, Equation \ref{eq:ImperfectArea} would become:
\begin{equation} \label{eq:realisticdiffuse}
    \frac{A_{D}}{A_{PM}} \leq \left[ \left( \frac{1}{\Omega_{PSF}} \ \frac{\pi}{\rho_d} \ \frac{F_p}{F_*} \right)^{-1} +1 \right]^{-1} \ .
\end{equation}
Using this treatment instead in the purely diffuse case, the allowed defective mirror area would be 99.99\% of the total mirror area.} We therefore conclude that any defective or damaged mirror area that acts like a perfect diffuse Lambertian reflector would be inconsequential. 

However, defective mirror area will realistically present both a diffuse and specular component. In the opposite extreme where we assume the defective mirror area is perfectly specular, defective area will scatter the host star light into the planet's core in the image plane and the mission will have extremely low tolerance to mirror damage. In other words, the integral of the BRDF in Equation \ref{eq:nonsimplifiedBRDF} will approach 1, and the allowed defect fraction will become:
\begin{equation} \label{eq:PerfectSpecular}
    \frac{A_{D}}{A_{PM}} \leq \frac{F_p}{F_*} = 3.6 \revision{\times} 10^{-10} \ .
\end{equation}
In this scenario, the allowed fraction of defective area is equal to the ratio of the planet to star flux. For the Earth-Sun case in the V band (i.e., Equation \ref{eq:EarthSunVband}), the defective area would only need to be $\sim$3.6$\times$10$^{-10}$ times the area of the primary mirror before the stray light requirement we are considering is violated (e.g., before stray light from mirror defects exceeded the brightness of the Earth-like planet within its PSF area). For reference, on a primary mirror with 6 m diameter, this would be $\sim$0.01 mm$^{2}$ of damage. \revision{We would like to stress here that a case where damaged area is a perfect specular reflector is fundamentally unrealistic. For an idealized mirror surface assumed to be a perfect specular reflector, light from the host star is coherent and can be removed by a coronagraph. Thus, if damaged area remains a perfect specular reflector, then the stray light should still be coherent and removable. In this example, we assumed that all stellar light reflected by the damaged area was incident into the planet's PSF, and this was meant purely to give the reader an intuitive sense on the difference between diffuse and specular reflection.}

Clearly the allowed defective mirror area may vary broadly depending on the assumed contribution from diffuse and specular reflection. To explain the effects of stray light, we must therefore consider more realistic BRDF models for a damaged mirror.

\subsection{Peterson Model \& Micrometeoroid Damage} \label{subsubsec:PetersonModel}

To model the complex scattering behavior of craters from micrometeoroid impacts, we will adopt the Peterson BRDF model \citep{Peterson2012brdf}, an idealized model for mirror digs, \revision{which was previously validated against the FRED optical software \cite{FREDsoftware} as a model of micrometeoroid damage for the LISA mission \cite{Khodnevych2020LISAMeteor}}. The Peterson BRDF for digs is isotropic and given by:
\begin{equation} \label{eq:PetersonBRDF}
    f_{r}(\theta_i, \theta_s) = \frac{N_D D_{mm}^{2}}{4} \left[ 1 + \frac{\pi^{2} D_{mm}^{2}}{4 \lambda^{2}} \left( 1 + \frac{(\Vec{V_{r}} - \Vec{V_{ps}})^{2}}{l_D^{2}} \right)^{-3/2} \right] \ ,
\end{equation}
\begin{wrapfigure}[12]{r}{0.5\textwidth}
\vspace{-25pt}
  \begin{center}
    \includegraphics[width=0.49\textwidth]{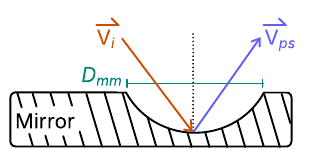}
  \end{center}
  \vspace{-15pt}
  \caption{\revision{An example of a damage crater (represented by a `dig' in the Peterson \citep{Peterson2012brdf} model) with damage diameter $D_{mm}$. For incoming light with a direction of incidence of $\Vec{V_{i}}$, the direction of perfect specular reflection is given by $\Vec{V_{ps}}$. The direction of reflectance would be determined by the direction over which you are interested in quantifying the amount of reflected light. }}
  \label{fig:PetersonModel}
\end{wrapfigure}
where $D_{mm}$ is the typical crater diameter, $\Vec{V_{r}}$ and $\Vec{V_{ps}}$ are the vector directions of reflection and perfect specular reflection, $N_{D}$ is the number of craters per unit area of diameter $D_{mm}$, $\lambda$ is the wavelength of interest, and \revision{$l_D$ is the roll-off angle \cite[equation 20 from][]{Peterson2012brdf} given by}:
\begin{equation} \label{eq:PetersonRolloff}
    l_D = \left( \frac{4}{\pi^{4}} \right)^{1/3} \frac{\lambda}{D_{mm}} \ .
\end{equation}
\revision{Figure \ref{fig:PetersonModel} shows a simple diagram to visualize an impact crater on a mirror surface, and how $\Vec{V_{ps}}$ is defined.}
For this estimation we will consider craters from impactors with masses of $10^{-18}$ -- 1 gram. We will then convert impactor mass to crater diameter using optimistic and pessimistic models: Watts \cite{wattsModel1993,heaney1993} and McHugh and Richardson (MR) \cite{spenvis2024,spenviswebsite}, respectively. \revision{Here, the Watts model is more 'optimistic' for our purposes because it generally predicts smaller craters than the MR model for the same impactor, making the MR model more 'pessimistic'}. When integrating the Peterson BRDF, we use numerical integration with Python's SciPy package \citep{SciPy2020}. Again, these results are under the scenario where the host star is the sole stray light source and our mirror is a point source ($\theta_i$ = $\theta_r$ = 0) in the V-band ($\lambda$ = 0.545$\mu$m). The Watts model is given by:
\begin{equation} \label{Eq:watts}
    d_{c} = (\frac{\rho_{p}}{\rho_{t}})^{\frac{1}{3}}\, v^{\frac{2}{3}}\, d_{p} \ ,
\end{equation}
and the MR model is given by:
\begin{equation} \label{Eq:MR}
    d_{c} = 12.8\,  d_{p}^{1.2}\, v^{\frac{2}{3}}\, \rho_{t}^{\frac{1}{2}}\,  \cos^{\frac{2}{3}}(\alpha) \ ,
\end{equation}
in both cases $d_{p}$ is impactor diameter (cm), $v$ is impact velocity $[\frac{km}{s}]$ , $\rho_{t}$ is the target density [$\frac{g}{cm^3}$], $\alpha$ is the impact angle, and $\rho_{p}$ is impactor density. In both Equation \ref{Eq:MR} and \ref{Eq:watts}, impactor diameter is found from impactor mass, $m$, [g]:
\begin{equation} \label{Eq:dRho}
     d_{p} = 2\left(\frac{3 m}{4  \rho_{p}  \pi}\right)^{1/3} \ .
\end{equation}
To simplify our calculations, we assume all impactors are spherical particles with a density of \revision{2.5 g/cm$^{3}$ \cite{Love1994meteordensity,Mcdonnell1998meteoroid,Kikwaya2011meteordensity}}, the mirror is taken to have a density of glass of 2.7 g/cm$^{3}$ \cite{schaferModel2001}, impact velocities are held constant throughout this study at 20 km/s \cite{courpalais1985,grunModel} so that impact energies increase with mass, impacts occur normal to the surface, and there is no overlap between impact craters. We stress these calculations are meant to be a reasonable estimate of micrometeoroid damage.

\begin{figure}
    \centering
    \includegraphics[width=0.8\linewidth]{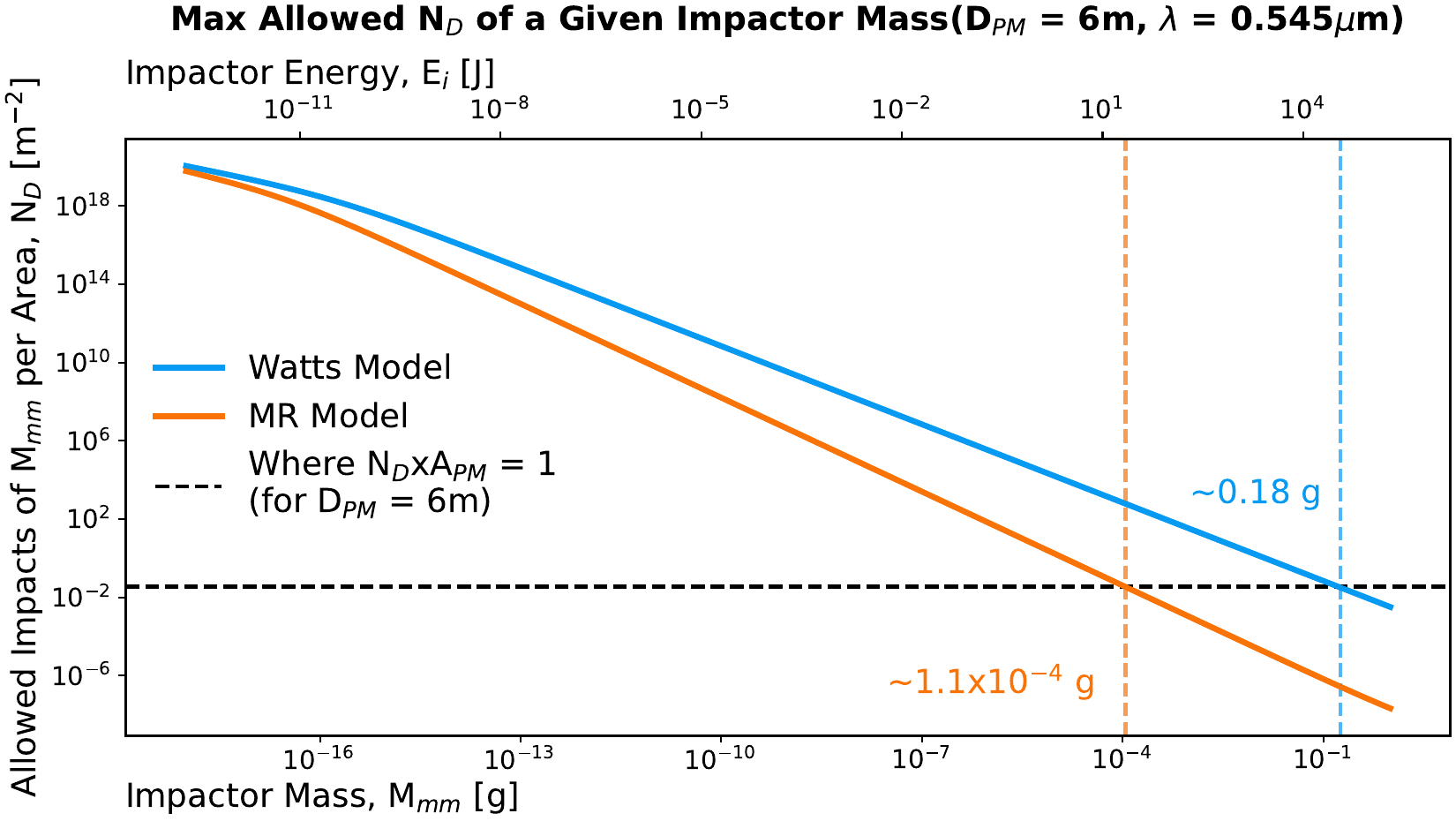}
    \caption{The maximum allowed number of impacts per unit area (N$_{D}$, m$^{-2}$) of a given impactor mass (M$_{mm}$, g) with energy E$_{i}$ (assuming a constant impact velocity of 20 km/s) to comply with the stray light requirement (Equation \ref{eq:nonsimplifiedBRDF}) under the assumption that the host star is the sole source of stray light for an Earth-Sun twin. We assume a V band wavelength ($\lambda$ = 0.545$\mu$m) and a primary mirror with a diameter of 6 meters. An optimistic \citep[Watts,][]{wattsModel1993,heaney1993} and pessimistic \citep[McHugh and Richardson, or MR,][]{spenvis2024,spenviswebsite} model relating impactor mass (M$_{mm}$) to crater diameter (D$_{mm}$) is used to determine the allowed number of impacts. For the MR and Watts models, impactor masses $\gtrsim$1.1$\times$10$^{-4}$ grams (22 J) and $\gtrsim$0.18 grams (36 kJ), respectively, would violate the stray light requirement after a single hit.}
    \label{fig:petersonestimate}
\end{figure}

Using the Peterson model, we find that a moderate amount of damage (e.g., resulting from a 10$^{-4}$ -- 10$^{-1}$g impactor mass) may greatly increase stray light, and that this number will strongly depend on the model adopted to relate impactor mass to crater diameter. Figure \ref{fig:petersonestimate} shows the maximum allowed number of impacts per unit area (N$_{D}$, m$^{-2}$) of a given impactor mass (M$_{mm}$) or energy (E$_{i}$) when using either the Watts or MR model to relate impactor mass to crater diameter (D$_{mm}$) following the stray light requirement of Equation \ref{eq:nonsimplifiedBRDF}. The black dashed line on Figure \ref{fig:petersonestimate} indicates where the allowed number of impacts, N$_{D}$, times the primary mirror area of a D$_{PM}$ = 6 m mirror would equal 1, i.e., a single impact would cause a violation of this stray light requirement. \revision{This single impact violation corresponds to a mass range of about three orders of magnitude between the two models ($\sim$1.1$\times$10$^{-4}$ g, 22 J for the MR model versus 0.18 g, 3600 J for the Watts model, taking our assumed impact velocity of 20 km/s) and reflects the large difference in estimated impact diameter with  micrometeor mass or energy between these two models. For comparison, the high energy micrometeoroid impact event that affected JWST mirror segment C3 in May 2022 was estimated to have an impact energy between 7 -- 26 J \citep{MenzelC3Energy,Menzel2023jwstreport}. So a single, similar impact at the high end of the impact energy estimate (22 J) would either cause a violation of our simple stray light  requirement or would be negligible depending on the model.} \revision{Furthermore, Figure \ref{fig:petersonestimate} demonstrates that the mission's sensitivity to micrometeoroid damage is a steep function of the impact energy for single hit events, therefore the few stochastic, lower probability impacts from high energy impactors will pose a significant stray light threat.}



In this section, we have demonstrated that stray light effects due to micrometeoroid damage on primary mirror area are not well understood, may vary broadly, and could jeopardize the success of an HWO-like mission. Considering this stray light source will be present in tandem with other challenges such as mirror roughness, intrinsic coating scatter, and contamination, HWO may require a procedure for mitigating the effects of infrequent large micrometeoroid impacting events. In the remainder of this study we add spatial-resolution to improve the fidelity of our micrometeoroid damage and stray light models, and explore the impact of micrometeoroid damage on mission science yield.


\subsection{Generalized Approach to Modeling Stray Light} \label{subsec:generalized}


To assess stray light more realistically, we aim to model a telescope mirror and a true sky background that are both spatially resolved. Until now, we have assumed the host star is the sole source of stray light while treating the primary mirror as a point source surface (where all primary mirror area is located at the center point of the mirror). To achieve a realistic stray light calculation that depends on pointing direction and telescope design, we must adjust our equations to find the stray light contribution from reflection off of $N$ number of points (of area $dA$) on a spatially resolved mirror. Furthermore, we will scale up to assuming additional sources of stray light at varying separations from the pointing direction, which will eventually allow us to use a map of the true galactic background. 


\subsubsection{Spatially Resolved Primary Mirror} \label{subsubsec:SpatialPM}

We now consider the scenario where the host star is still the sole source of stray light, but instead of treating the primary mirror as a point source, we consider a spatially resolved surface. In this scenario, the stray light flux can be written as:
\begin{equation} \label{eq:initialSpatial}
    F_{SL} = \int_{A} \left[ F_* \ A_{D} \ cos(\theta_{i,0,dA}) \ \int_{\omega_r} f_r(\theta_{i,0,dA}, \phi_{i,0,dA}, \theta_{r}, \phi_{r}) \ cos(\theta_r) \ d\omega_r \right] dA\ ,
\end{equation}
which can be approximated as a sum over a finite number of dA elements if we divide the mirror into $N_{dA}$ points of equal area across the mirror surface:
\begin{equation} \label{eq:initialSpatial}
    F_{SL} = \sum_{N_{dA}} \left[ F_* \ A_{D,dA} \ cos(\theta_{i,0,dA}) \ \int_{\omega_r} f_r(\theta_{i,0,dA}, \phi_{i,0,dA}, \theta_{r}, \phi_{r}) \ cos(\theta_r) \ d\omega_r \right] \ .
\end{equation}
Here the flux of stray light (F$_{SL}$) is calculated as the sum of host star stray light contributions from the damaged area on each mirror point ($A_{D,dA}$). In this case, we use the subscript `$i,0$' to indicate light coming from the pointing direction and the subscript `$i,0,dA$' to indicate light coming from the pointing direction from the perspective of the area element $dA$. Note, as mirror points move away from the center, they have a non-zero reflectance angle ($\theta_r$) to reflect light into the secondary mirror; thus, each point has a normal vector oriented such that the light coming from the telescope's pointing direction ($\theta_{i,0}$ = $\phi_{i,0}$ = 0$^{o}$) will have perfect specular reflection into the secondary mirror. This means the angle of incidence from the perspective of each area element, $\theta_{i,0,dA}$, will change across the mirror surface. $F_{*}$ gives power per area and is thus constant; and we further assume that each mirror point across our surface has the same amount of area ($A_{dA}$ remains constant) and mirror defects are equally distributed across all mirror points (A$_{D,dA}$ remains constant).

\subsubsection{Spatially Resolved Sky Background}

Up until now we have only considered the host star as the sole source of stray light, in reality, there will be many stars near the pointing direction which can contribute stray light. Additionally, for a telescope with a barrel, mirror points across the surface will have an optical path to different areas on sky. Considering the sky area ($s$) that has an optical path to at least one point on the primary mirror, we can define our full stray light calculation:
\begin{multline} \label{eq:FullSLRequire}
    F_{SL} = \int_{s} \ \int_{A} \left[ F_{s} \ A_{D,dA} \ cos(\theta_{i,s,dA}) \ \int_{\omega_r} f_r(\theta_{i,s,dA}, \phi_{i,s,dA}, \theta_{r}, \phi_{r}) \ cos(\theta_r) \ d\omega_r \right] dA \ ds\ ,
\end{multline}
which can be approximated as a double sum if we split the sky area into $N_s$ evenly spaced gridpoints:
\begin{multline} \label{eq:FullSLRequire}
    F_{SL} = \sum_{N_{s}} \ \sum_{N_{dA}} \left[ F_{s} \ A_{D,dA} \ cos(\theta_{i,s,dA}) \ \int_{\omega_r} f_r(\theta_{i,s,dA}, \phi_{i,s,dA}, \theta_{r}, \phi_{r}) \ cos(\theta_r) \ d\omega_r \right] \ .
\end{multline}
This gives the sum of the stray light contributions from the flux originating from point $s$ on the sky incident on the damaged area on each mirror point ($A_{D,dA}$) across all sky points $N_s$. Here, $\theta_{i,s,dA}$ refers to the angle of incidence from sky point $s$ in area point $dA$'s perspective frame. 


In the following section (\S\ref{sec:modelmethods}), we describe additional relevant computational methods for calculating realistic stray light estimates before presenting results of our work in \S\ref{sec:Results}.

\section{Modeling Methods} \label{sec:modelmethods}

To account for the sources of stray light in a realistic observation, some of our simulation results use compiled maps of the galactic background from Gaia in a specific bandpass (\S\ref{subsec:GaiaMap}). The area of the sky visible to the telescope, or the stray light background, ultimately depends on the pointing direction, the ratio of the length of the telescope barrel to primary mirror diameter, the relative asymmetry of the barrel (i.e., for a scarfed barrel) and barrel pointing orientation (i.e., roll, in the case of an asymmetrical or scarfed barrel). We briefly describe the model developed to simulate stray light observations in \S\ref{subsec:FullSLModel}, but save more detailed descriptions for Appendix \ref{sec:FullSLModelDescriptionAppendix}.
Finally, we will consider basic estimates of potential micrometeor damage rates to draw conclusions on the importance of stray light considerations for an HWO-like mission, the method for determining these damage approximations is described in \S\ref{subsec:MeteorDamModel}.

\subsection{Gaia Background Flux Mapping} \label{subsec:GaiaMap}


For a particular passband, we created sky background maps by finding the cumulative flux per degree$^{2}$ bin across all galactic latitudes and longitudes using the entire Gaia data release 3 (DR3) archive \citep{Prusti2016gaiamission,Vallenari2023gaiaDR3}. We present flux maps in Gaia G band, Johnson's V band, and Cousin's Ic band. The stray light found will change based on the passband used, as the flux map will be different and the BRDF integrations have a dependence on wavelength, though we will present results from the V band in this study as it is most relevant to HWO \citep{Stark2024yield}. For the Gaia G band flux, values are directly reported in Photons/m$^{2}$/s from the Gaia DR3 archive, thus we sum this quantity directly per sky area bin. However, to find the flux maps in Johnson's V and Cousin's Ic bands, we convert the Gaia G band magnitude to a V or Ic band magnitude and then find the corresponding flux in W/m$^{2}$/$\mu$m. Conversions from Gaia G magnitude to Johnson-Cousins photometric passband require BP-RP color information on the source and use the following transformation polynomial:
\begin{equation} \label{Eq:ToVbandMag}
    G - M = A \ + \ B(G_{BP} - G_{RP}) \ + \ C (G_{BP} - G_{RP})^{2} \ + \ D (G_{BP} - G_{RP})^{3} \ ,
\end{equation}
where transformation to a given magnitude, $M$ (Johnson's V or Cousin's Ic), is described with the transformation polynomial coefficients ($A$, $B$, $C$, and $D$) given in Table \ref{tab:PolyCoeff} from \cite{Riello2021gaiaEDR3PhotometricTrans} (also reported in Table 5.9 of the Gaia DR3 Documentation\footnote{https://gea.esac.esa.int/archive/documentation/GDR3/}). The apparent magnitude can then be converted into a flux at that passband ($F_{M}$) via:
\begin{equation}
    F_{M} = F_{zp} 10^{-M/2.5} \ ,
\end{equation}
where $F_{zp}$ is the Vega zeropoint flux at that passband (from \cite{Bessell1998uvb} for Johnson's V and \cite{Bessell1979cousinsri} for Cousin's Ic), which are also given in Table \ref{tab:PolyCoeff}. 

\begin{table}[]
    \centering
    \setlength\extrarowheight{2pt}
    \begin{tabular}{|c|c|c|c|c|c|c|}
         \cline{3-6}
         \multicolumn{2}{c|}{ } & \multicolumn{4}{c|}{Transformation Polynomial Coefficients} & \multicolumn{1}{c}{ } \\
         \cline{3-7}
         \multicolumn{2}{c|}{ } & A & B & C & D & F$_{zp}$ [Wm$^{-2}$$\mu$m$^{-1}$] \\
         \hline
         \multirow{2}{*}{M} & V & -0.02704 & 0.01424 & -0.2156 & 0.01426 & $3.6\revision{\times}10^{-8}$ \citep{Bessell1998uvb} \\
         \cline{2-7}
          & Ic & 0.01753 & 0.76 & -0.0991 & 0 & $1.2\revision{\times}10^{-8}$ \citep{Bessell1979cousinsri} \\
          \hline
    \end{tabular}
    \vspace{1mm}
    \caption{The transformation polynomial coefficients ($A$, $B$, $C$, and $D$) from \cite{Riello2021gaiaEDR3PhotometricTrans} for converting a Gaia G band magnitude to either Johnson's V or Cousin's Ic band using equation \ref{Eq:ToVbandMag}, and the Vega zeropoint flux for Johnson's V \citep{Bessell1998uvb} and Cousin's Ic \citep{Bessell1979cousinsri}.}
    \label{tab:PolyCoeff}
\end{table}

\begin{figure}
    \centering
    \includegraphics[width=\linewidth]{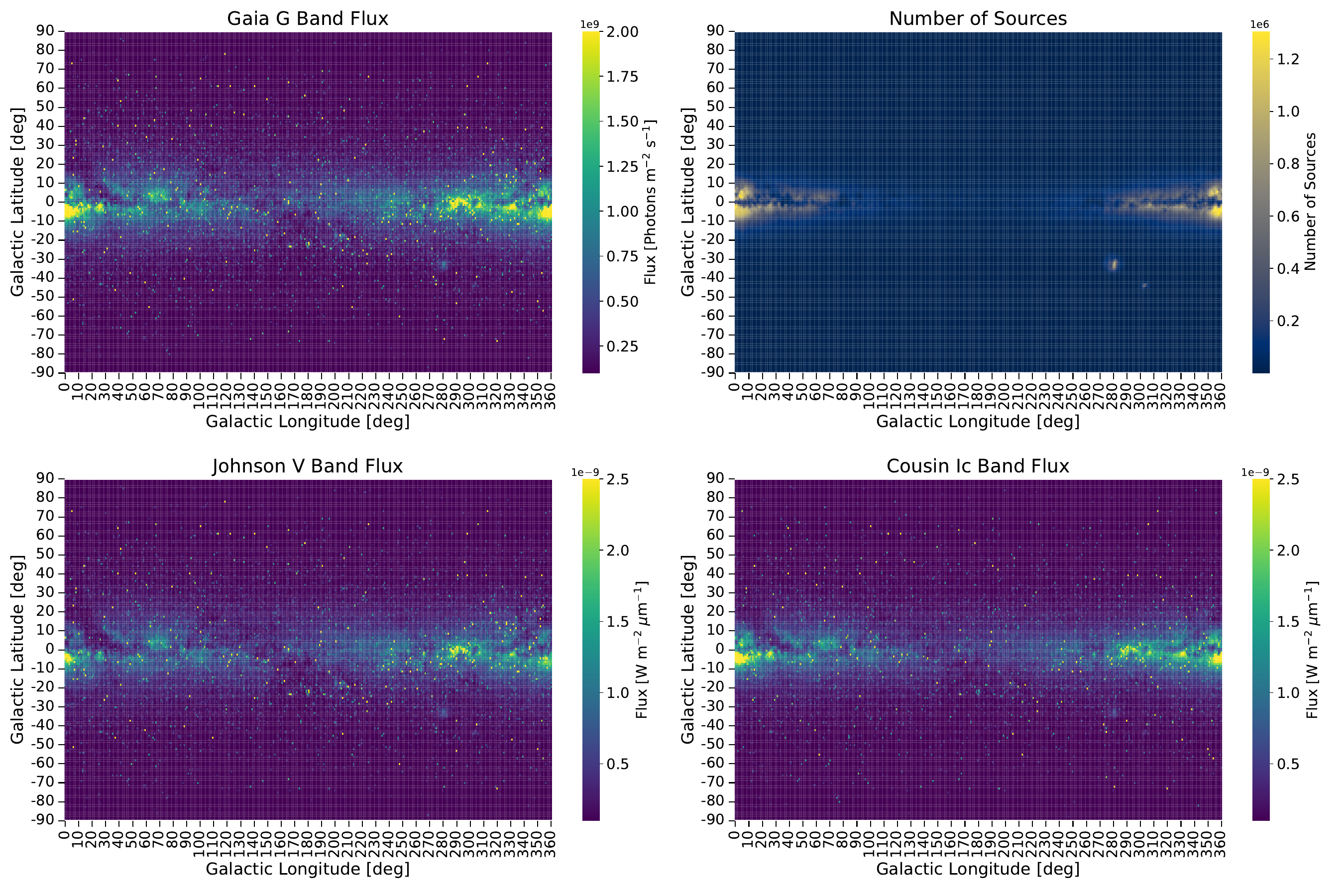}
    \caption{Cumulative incident flux from the sky background in Gaia G (top left, Photons m$^{-2}$ s$^{-1}$), Johnson's V (bottom left, W m$^{-2}$ $\mu$m$^{-1}$), and Cousin's Ic band (bottom right, W m$^{-2}$ $\mu$m$^{-1}$). Additionally shown is the total number of sources per bin (top right). These maps were created using the entire Gaia DR3 archive \citep{Prusti2016gaiamission,Vallenari2023gaiaDR3}; conversions from Gaia G band magnitudes to Johnson's V and Cousin's Ic bands used photometric transformations defined by \cite{Riello2021gaiaEDR3PhotometricTrans} and zeropoint fluxes from \cite{Bessell1979cousinsri,Bessell1998uvb}.}
    \label{fig:FluxMaps}
\end{figure}

Figure \ref{fig:FluxMaps} shows the flux maps in all 3 passbands presented here (Gaia G, Johnson's V, and Cousin's Ic), as well as a map of the total number of sources in each bin. These maps were created by running a Gaia query to sum the fluxes of all sources in each degree$^{2}$ bin (first converting to the desired passband, if necessary).

\subsection{Simulating a Stray Light Observation} \label{subsec:FullSLModel}

For a given target and set of user inputs (e.g., mirror size, barrel dimensions, BRDF to use, etc), our model estimates the stray light contribution considering every mirror area element and every accessible background sky point.
The observational setup includes the pointing direction, primary mirror size, telescope barrel design, orientation of the barrel with respect to the pointing direction (i.e., the roll angle), the resolution of the primary mirror model (i.e., the number of mirror area points to use), and the desired passband; unless otherwise specified, we will present results from the V band ($\lambda$ = 0.545$\mu$m). 


For the sky area around the pointing direction, the model will loop through each sky point (N$_{s}$) and calculate its stray light contribution from the reflection off every mirror point (N$_{dA}$). In other words, the model evaluates the summation of the integrated BRDF model across all mirror points for every sky point (Equation \ref{eq:FullSLRequire}). One exception to this general approach is for sources extremely close to the pointing direction. For the sky grid box that contains the pointing direction, approximating the stray light as the cumulative flux from all sources in that box at the same pointing direction will overestimate stray light, as sources closer to the pointing direction contribute more heavily. Thus, for the gridbox containing the pointing direction, we resolve every source in the box and calculate their stray light contributions individually, which gives us an accurate estimation of the stray light near the pointing direction.

To simplify the bounds of integration when evaluating the \revision{radiance} into $\omega_{r}$, we evaluate integrals in the reference frame of the direction of reflection for a given mirror area point; this allows us to consistently integrate $\phi_{r}$ from 0 to 2$\pi$ and $\theta_{r}$ from 0 to $\lambda$/D. To prepare this calculation, the model defines all relevant directions for calculating the stray light contribution from a sky point on a mirror area point and then conducts 3D axes rotations to transform into the direction of reflection reference frame. We save the exact description of these transformations for Appendix \ref{sec:FullSLModelDescriptionAppendix}. The full stray light model developed in this work is publicly available on GitHub\footnote{https://github.com/mgialluca/StrayLightBackgroundModel}.

\subsection{Yield Calculations} \label{subsec:YieldMethods}

Here we explore the effects of stray light on exoplanet science yield. Science yield in this study is defined as the number of extrasolar habitable zone Earth-like planet candidates that may be discovered by HWO. To calculate yield for a given set of mission parameters, we adopt the Altruistic Yield Optimization (AYO) method developed by \cite{Stark2014AYO,Stark2019yieldandrevisits,Stark2024yield}. Previous AYO calculations included stray light originating from nearby binary star companions assuming a reasonable particle size distribution on the mirror to represent surface roughness and contamination \citep{Stark2019yieldandrevisits}, but did not account for stray light from any other source. In this work, we account for stray light originating from micrometeoroid damage on the primary mirror in addition to the binary star contribution from surface contamination adopted in previous studies \citep{Stark2019yieldandrevisits,Stark2024yield}.


To conduct our yield calculations, we adopt all baseline astrophysical and coronagraph inputs from \cite{Stark2024yield} for their "Scenario A" and "Scenario F" cases. Both scenarios are based off the LUVOIR-B baseline design, but adjusted to a 6 m inscribed diameter. Scenario A is very similar to the baseline design, with only one alteration to minimize the number of aluminum-coated mirrors in favor of silver, which will increase throughput. Scenario F represents a best-case design scenario by adopting the following design changes in addition to minimizing aluminum coatings: operating two V band coronagraphs in parallel, adopting a model-based PSF subtraction strategy, using an improved coronagraph detector design instead of the EMCCD detector used in the LUVOIR-B baseline case, and adopting a phase-induced amplitude apodizer coronagraph offering higher throughput than the LUVOIR-B baseline DM-assisted charge six vortex coronagraph \citep{Stark2024yield}. We refer the reader to \cite{Stark2024yield} for further information on the astrophysical and coronagaph inputs to AYO. Following \cite{Stark2024yield}, we use the Habitable Worlds Observatory Preliminary Input Catalog \citep[HPIC,][]{Tuchow2024hpic} to select target stars for observation, but restrict targets to be within 30 pc and brighter than magnitude 9 which allows the model to select from a pool of 1343 target stars. 




\subsection{Telescope Barrel Effects on Highest Mass Impactors} \label{subsec:MeteorDamModel}

\revision{To illustrate the impact of micrometeoroid damage, and potential steps to mitigate it, we calculate the micrometeoroid flux at the telescope's orbit and consider the shielding capabilities of three different telescope barrel designs based on HWO’s three Exploratory Analysis Cases (EACs). For each design, we define four damage cases corresponding to the largest expected impactor energy at a
with a 2.3\% probability (2$\sigma$ probability of a two-sided distribution) over 1 or 5 years, when pointed in the RAM or Anti-RAM directions (Table \ref{tab:MMImpactMasses}). 
The RAM direction is pointing in the direction of motion of the telescope (leading to higher energy impacts), while the Anti-RAM points opposite to the direction of motion (leading to lower energy impacts). To standardize our results for the barrel geometries explored in Table \ref{tab:MMImpactMasses}, we adopt
a fixed 6 meter inscribed diameter and adjust the barrel dimensions accordingly from the EAC
cases. The barrel design affects the highest probable impacts by changing the amount of the sky that is accessible to the telescope mirror (i.e., a longer barrel will reduce the sky area visible to the mirror, and lead to lower probable impact energies). NASA's MEM3 software was used because its limiting mass of a microgram was found to be smaller than the largest predicted impactor for the 3 EAC cases \cite{Moorhead2020replicate}.}

\revision{
To find the largest impactor, we model one orbit of the observatory placed at L2 in AGI's Systems Tool Kit (STK) \cite{stk} and feed this into NASA's MEM3 software \cite{moorhead2020nasa} to constrain micrometeoroid mass flux. 
Specifically, MEM3 provides the cumulative flux of micrometeoroid masses greater than or equal to a microgram in every 1$^\circ$$\times$1$^\circ$ sky bin across 1 km/s wide velocity bins, averaged over one orbit. We then use the stray light model (see Apendix \ref{sec:FullSLModelDescriptionAppendix}) to divide the area of sky that can access the primary mirror into grid points based on the barrel design, and attribute each point a value of 0 to 1 defining the fraction of the mirror with a clear optical path to that point, which we also refer to as the mirror coverage fraction (MCF).  
For each velocity bin ($v$), we sum the MEM3 mass flux ($F_{j,v}$) multiplied by the mirror area ($A_{PM}$) and respective MCF ($MCF_{j}$) over all sky grid points ($j$) to get the total rate of impacts per unit time as a function of impactor mass, $m$. 
The total number of impacts of mass $m$ per unit time as a function of velocity ($N_v$) can then be found by scaling this summation by the Gr{\"u}n flux \cite{grunModel} of impacts of mass $m$ ($F_{G}(m)$) divided by the Gr{\"u}n flux of MEM3's limiting mass ($F_{G}(10^{-6}g)$) \cite[taken from][]{moorhead2020nasa}:
\begin{equation} \label{Eq:mmScale}
        N_{v} = \sum_{j} F_{j,v} \ MCF_{j}  \ A_{PM} \ \frac{F_G(m)}{F_G(10^{-6}g)} \ .
\end{equation}
The flux values reported by MEM3 are scaled from the Gr{\"u}n flux and include all masses larger than a given limiting mass. Therefore equation \ref{Eq:mmScale} can be used to rescale the reported flux to a new limiting mass.}

\revision{
In MEM3, the velocity that produces the highest $N_v$ across all masses for RAM direction pointing is 55.5 km/s, and for the Anti-RAM direction it is 8.5 km/s. In either pointing direction, we look for the largest mass which produces an $N_v$ corresponding to a probability of 2.3\% (2$\sigma$) that 1 impact will occur in a given time period. This is found using a Poisson distribution: 
\begin{equation} \label{Eq:poisson}
        \mathbb{P}(2.3\%) =  \lambda t \, \times \, e^{-\lambda t}  \ ,
\end{equation}
 where $\lambda$ is the $N_v$ corresponding to a single hit impact probability of 2.3\% across a time period $t$.
 }

\begin{table}[]
    \centering
    \begin{tabular}{|c|c|c|c|c|c|c|}
       \hline
        \rule{0pt}{20pt} & \multicolumn{2}{|c|}{Barrel Geometry} & \multicolumn{4}{|c|}{Largest Single Hit Impact Energy [J] to 2$\sigma$ Probability} \\
       \hline
       \rule{0pt}{20pt} Case ID & \makecell{Long\\Side [m]} & \makecell{Short\\Side [m]} & RAM 1 yr & RAM 5 yr & Anti-RAM 1 yr & Anti-RAM 5 yr \\
       \hline
       \hline
       1 & 19.2 & 9.7 & 10.75 & 45.74 & 0.05 & 0.27 \\
       \hline
       2 & 14.5 & 10.3 & 13.46 & 56.52 & 0.07 & 0.32 \\
       \hline
       3 & 13.6 & 7.1 & 18.94 & 77.78 & 0.10 & 0.47 \\
       \hline
    \end{tabular}
    \vspace{1mm}
    \caption{The barrel dimensions for the 3 telescope design cases we consider here, and the maximum micrometeoroid impactor energies [J] found for each barrel design. All barrel geometries use an asymmetric barrel and adopt a 6 m inscribed primary mirror diameter. The long and short sides are the length of each side of the telescope barrel. For each case, we list the largest single hit expected to 2$\sigma$ (2.3\%) probability when the telescope is assumed to point in either the RAM or Anti-RAM direction for 1 or 5 years.}
    \label{tab:MMImpactMasses}
\end{table}

\section{Results} \label{sec:Results}

Here we present the results of the stray light calculations. 
We first compare the stray light predicted when using the fully resolved sky background versus the host star only (\S \ref{subsec:comparefullmodelresults}), then we quantify the reduction in expected exoEarth yield for various impactor sizes (\S \ref{subsec:ScienceYieldResults}).


\subsection{Comparing Host Star Stray Light to a Fully Resolved Sky Background} \label{subsec:comparefullmodelresults}

\begin{figure}
    \centering
    \includegraphics[width=0.98\linewidth]{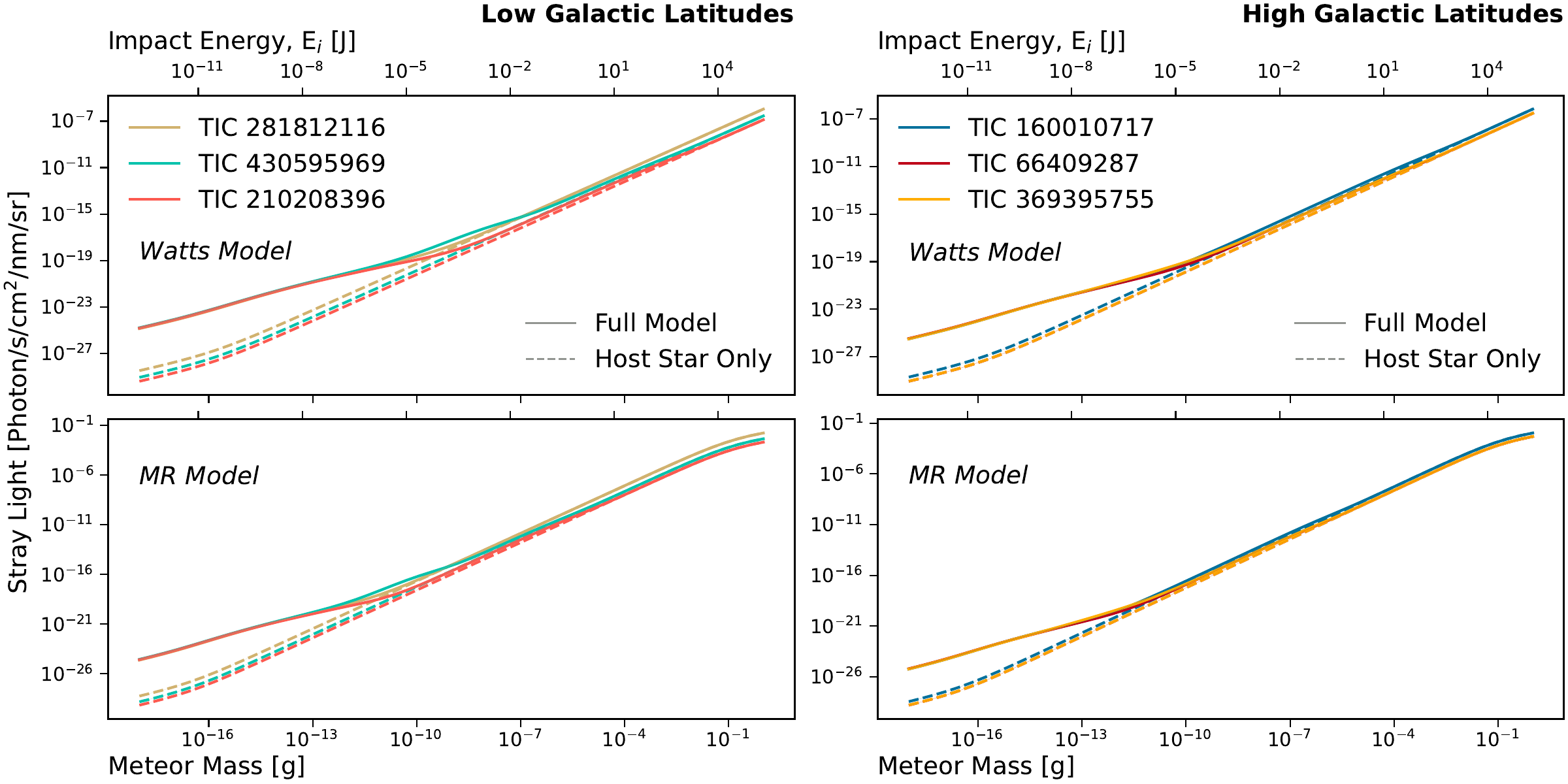}
    \caption{A comparison of the stray light found when considering all sources in a target's resolved sky background (solid lines) versus only the host star (dashed lines) as a function of the micrometeoroid mass (or impact energy) for a single-hit event. From the truncated HPIC catalog \citep{Tuchow2024hpic} used in this study, the 3 targets at the lowest and highest galactic latitudes are shown in the left and right columns, respectively. The top and bottom rows demonstrate the Watts and MR models, respectively, for relating impactor mass/energy to crater diameter. In all cases, considering all sources in the resolved sky background only notably increases the computed stray light for single-hit impactors of mass $\lesssim$10$^{-10}$ or impact energies $\lesssim$10$^{-5}$ J. }
    \label{fig:FullComparetoHostStar}
\end{figure}

Given the specular nature of the Peterson BRDF model, we find that the additional contribution of stray light from the full sky background is negligible compared to the stray light from the host star for all but the smallest impacting events (less than 2 mJ). Figure \ref{fig:FullComparetoHostStar} shows a comparison of the stray light model using the sources in the full resolved sky background from Gaia versus just the target host star, selected from the HPIC list. To determine if there is any dependence of stray light on the proximity of a target's distance to the galactic disk, we test the three sources with the lowest (left subplots) and highest (right subplots) galactic latitudes in the HPIC catalog \citep[][including the truncation due to distance and magnitude cuts described in \S\ref{subsec:YieldMethods}]{Tuchow2024hpic}. In the figure, the top and bottom rows use the Watts and MR models, respectively, to relate impactor mass/energy to crater diameter. When using the full sky background, assumed barrel geometry will have an effect on the calculated stray light. This is not the case when considering the host star only, as every point on the primary mirror will have an optical path to the host star in the pointing direction for any barrel geometry. For this result shown in Figure \ref{fig:FullComparetoHostStar}, we use the smallest barrel considered in Table \ref{tab:MMImpactMasses} (13.6 $\times$ 7.1 m) when using the full sky background, as this would maximize the resulting stray light.

Demonstrated by Figure \ref{fig:FullComparetoHostStar} in all cases, considering the full sky background of an observation can increase overall stray light by up to 3 -- 4 orders of magnitude for impact energies less than $\sim$10$^{-5}$ J. 
For impact energies greater than $\sim$10$^{-5}$ J, considering stray light from the full sky background produces a negligible increase compared to the stray light from the host star only. Furthermore, while there is a small difference between low and high galactic latitudes, this difference is not large enough to warrant distinction between the two cases in later results. While an impact energy of 10$^{-5}$ J produces an increase in the scattered light from the resolved sky background over the host star only of 2 -- 30 times for low Galactic latitudes, which is markedly different than the 2 -- 6 times increase seen at high Galactic latitudes, by an impactor energy of 2 mJ and above, the difference between the two latitudes disappears. We finally note that, while lower impact energies can show great differences between the full sky background and the host star only case, the absolute stray light is still many orders of magnitude less than single events from higher energy impactors. However, we caution that the reflection model developed in this work is highly idealized and only considers single-scattering. 

Based on the result in this section (Figure \ref{fig:FullComparetoHostStar}), if yield is affected for impact energies less than or equal to 2 mJ, the full sky background must be resolved to account for the increase in stray light. However, if yield is only affected for impact energies greater than 2 mJ, the stray light from the full resolved sky background may be adequately approximated by the stray light from the host star only, as demonstrated by Figure \ref{fig:FullComparetoHostStar}.

\subsection{The Effect of Stray Light from Micrometeoroid Damage on Science Yield} \label{subsec:ScienceYieldResults}

\begin{figure}
    \centering
    \includegraphics[width=0.98\textwidth]{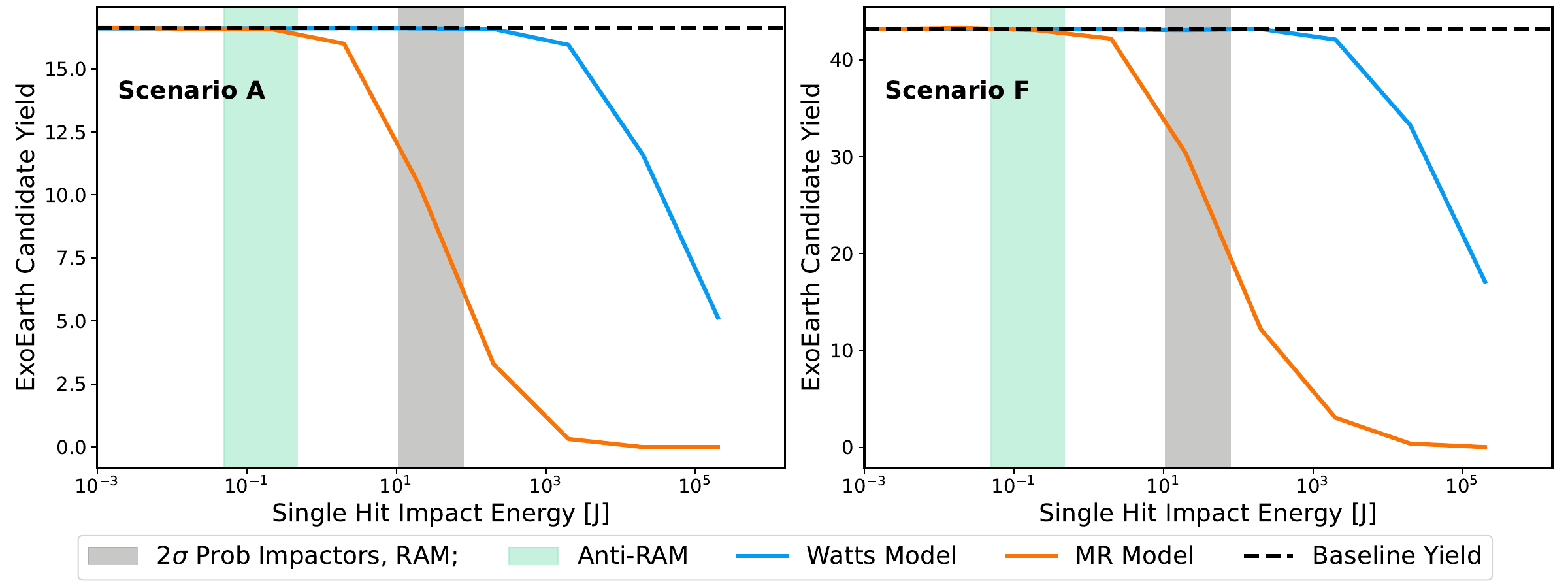}
    \caption{The exoEarth candidate yield over a 2 year survey calculated by AYO versus the impactor energy from a single-hit micrometeroid strike on the primary mirror. The Watts and MR models relating impactor mass/energy to crater diameter were tested, and are shown by the blue and orange lines, respectively. The left and right panels adopt the Scenario A and F telescope designs, respectively, from \cite{Stark2024yield}\revision{; note the changing y-axis scale for the separate scenarios}. The horizontal dashed black line in both panels gives the baseline yield, found when assuming there is no stray light contribution from micrometeoroid damage. The largest 2$\sigma$ probability impact cases shown in Table \ref{tab:MMImpactMasses} \revision{for RAM or Anti-RAM direction pointing fall within the grey and teal shaded regions, respectively.} If the more pessimistic MR model relating impactor mass to crater diameter is adopted, the impact cases identified \revision{for RAM direction pointing} would lead to a reduction in yield of 30 -- 60\%.}
    \label{fig:YieldCurves}
\end{figure}

Here we present the yield estimated by AYO \citep{Stark2014AYO,Stark2019yieldandrevisits,Stark2024yield} for 2 years of observation. We calculate the stray light contributed by a single-hit case with an array of impactor energies from 10$^{-13}$ J -- 200 kJ; each impact case is assumed to affect the full 2 years of observation. Figure \ref{fig:YieldCurves} shows the relationship between exoEarth candidate yield and impactor masses for a single-micrometeoroid hit case testing both the optimistic Watts and pessimistic MR models relating impactor mass to crater diameter. The left and right panels of Figure \ref{fig:YieldCurves} adopt the Scenario A and F coronagraph designs from \cite{Stark2024yield}, respectively, roughly representing the range of mission performance scenarios possible for HWO. The largest single impactor masses predicted to 2$\sigma$ probability from Table \ref{tab:MMImpactMasses} are indicated in Figure \ref{fig:YieldCurves} by the vertical grey shaded region \revision{for RAM direction pointing, and the light teal shaded region for Anti-RAM direction pointing}. The MR and Watts crater diameter models predict that single-hit events with impactor masses less than or equal to 10$^{-6}$ g (0.2 J) and 10$^{-3}$ g (200 J) will have negligible effect on the science yield, respectively. As discussed in the previous section, stray light from the fully resolved sky background can be reasonably approximated by considering only the host star for impacts above 2 mJ. Since science yield is only affected by impactors with energies $\geq$0.2 J, the results presented here compute stray light from the host star alone when modeling all sources in the truncated HPIC catalog \citep{Tuchow2024hpic} to reduce computational costs.

In addition to Figure \ref{fig:YieldCurves}, Table \ref{tab:yieldresults} gives the inferred percent loss in yield from each of the impact cases tested for both scenarios A ("Scen. A") and F ("Scen. F"). In either scenario, yield begins to be affected for impact energies of 2 or 2000 J for the MR and Watts models, respectively. For comparison, the JWST C3 May 2022 impact was found to have an impact energy between 7 -- 26 J \citep{MenzelC3Energy,Menzel2023jwstreport}.


If the MR model is assumed, which predicts larger craters than the Watts model for the same impact energy, yield may be reduced by 30 -- 60\% for the largest single-hit impact events predicted to 2$\sigma$ probability \revision{for RAM-direction pointing} after 1 and 5 years (from Table \ref{tab:MMImpactMasses}). Conversely, if the Watts model is assumed, none of the cases identified in Table \ref{tab:MMImpactMasses} would have a significant effect on yield. As the potential implications for reduction in science yield may range from negligible to extreme, this suggests the relationship between impact energy and crater diameter must be studied in much greater detail early in the mission design phase. Additionally, greater knowledge of the expected micrometeoroid population may be greatly beneficial to defining design requirements based on stray light. 

\begin{table}[]
    \centering
    \begin{tabular}{|c|c|c|c|c|c|c|c|c|c|}
       \hline
       \rule{0pt}{15pt} & & Mass [g] & $\leq$10$^{-6}$ & 10$^{-5}$ & 10$^{-4}$ & 10$^{-3}$ & 0.01 & 0.1 & 1 \\
       \hline
       \rule{0pt}{15pt} & & Energy [J] & $\leq$0.2 & 2 & 20 & 200 & 2$\times$10$^{3}$ & 2$\times$10$^{4}$ & 2$\times$10$^{5}$ \\
       \Xcline{1-10}{1.2pt}
       \rule{0pt}{25pt}\multirow{4}{*}{\rotatebox[origin=c]{270}{\makecell{Yield\\Reduction [\% loss]}}} & \multirow{2}{*}{\rotatebox[origin=c]{270}{Scen. A}} & Watts & $<$1\% & $<$1\% & $<$1\% & $<$1\% & 3.95\% & 30.3\% & 69.1\% \\
       \cline{3-10}
       \rule{0pt}{25pt} & & MR & $<$1\% & 3.68\% & 37.2\% & 80.2\% & 98.1\% & 100\% & 100\% \\
       \Xcline{2-10}{1.2pt}
       \rule{0pt}{25pt} & \multirow{2}{*}{\rotatebox[origin=c]{270}{Scen. F}} & Watts & $<$1\% & $<$1\% & $<$1\% & $<$1\% & 2.46\% & 22.9\% & 60.4\% \\
       \cline{3-10}
       \rule{0pt}{25pt} & & MR & $<$1\% & 2.25\% & 29.7\% & 71.7\% & 93.0\% & 99.0\% & 99.9\% \\
       \Xcline{1-10}{1.2pt}
       
    \end{tabular}
    \vspace{3mm}
    \caption{The percentage in yield reduction from the baseline for all cases tested and shown in Figure \ref{fig:YieldCurves}.}
    \label{tab:yieldresults}
\end{table}








\section{Discussion} \label{sec:discussion}

\subsection{Initial Impact Testing} \label{subsec:ImpactTestDiscuss}

Recent laboratory testing has indicated that the Watts and MR models are bounding cases for the conversion of impactor mass/energy into crater diameter. In an initial study by Heliospace to characterize the effect of micrometeoroid impacts, it was found that an 8 J impact test on a ULE mirror blank resulted in a damage diameter of 14.11 mm (Joe Pitman, Heliospace, \textit{private comm.}). For comparison, using the Watts \citep{wattsModel1993} and MR \citep{spenvis2024,spenviswebsite} models used in this study for an 8 J impact would predict damage diameters of 2.2 mm and 24 mm, respectively. To produce a damage diameter of 14.11 mm, the Watts and MR models would require a 1940 J and 2.03 J impact, respectively.

\begin{figure}
    \centering
    \includegraphics[width=0.95\linewidth]{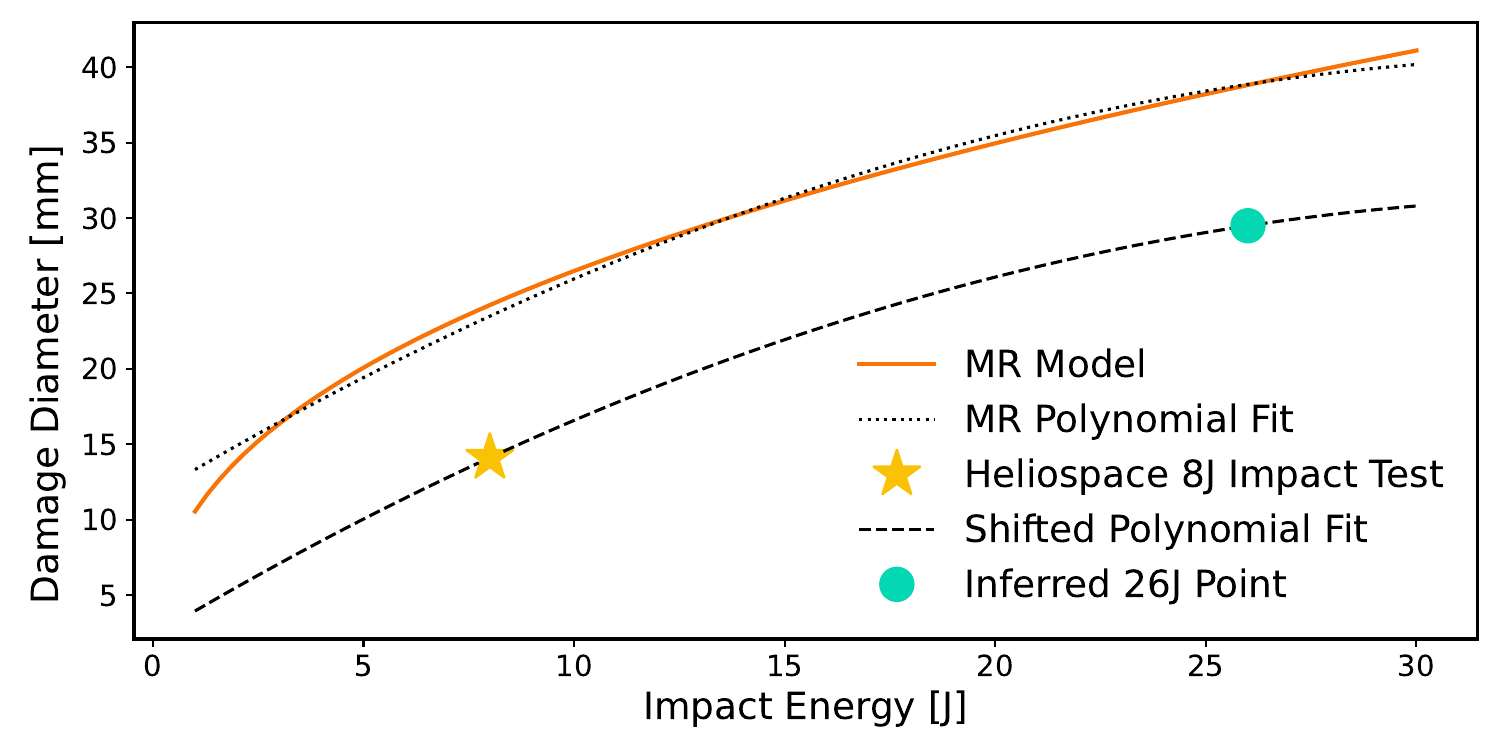}
    \caption{An illustration of the polynomial fitting process to infer a damage diameter for a 26 J impact based on the 8 J impact test from Heliospace. Damage diameter on the primary mirror [mm] is shown as a function of impact energy [J]. The original MR model is shown by the orange line and the polynomial fit by the black dotted line. The polynomial fit is then shifted down (black dashed line) to fit through the 8 J impact test by Heliospace (yellow star). This fit is then used to infer a damage diameter of 29.5 mm for a 26 J impact, shown by the aqua green circle.}
    \label{fig:polyfit}
\end{figure}

Our yield calculations can be used with the empirical data described above to estimate the maximum damage, and the corresponding yield reduction, for an impact of the magnitude of the JWST C3 May 2022 event. Inferred from Figure \ref{fig:YieldCurves}, a damage diameter of 14.11 mm would result in a loss in yield from the baseline of $\sim$4\% and $\sim$2\% for the Scenario A and F coronagraph models, respectively. As explained above, this level of damage is expected from an 8 J impact, which is believed to be around the energy of the C3 May 2022 impact event on JWST. Given that the estimated impact energy range from the C3 event is 7 -- 26 J \citep{MenzelC3Energy,Menzel2023jwstreport}, an impactor near the upper end of this range would likely produce a larger crater and an even greater reduction in yield. To estimate the effect of a 26 J impact, we can calibrate the MR model to the 14.11 mm impact test by fitting a 2nd order polynomial to the relationship between impact energy and crater diameter. We then translate this polynomial down to fit through the 8 J impact test from Heliospace. This relationship would lead to a damage diameter of 29.5 mm for a 26 J impact, the higher end of the range inferred from the C3 event. The polynomial fitting process is illustrated in Figure \ref{fig:polyfit} as a visual guide. Inferred from Figure \ref{fig:YieldCurves}, a damage diameter of 29.5 mm would result in a loss of yield from the baseline of $\sim$30\% in Scenario A and $\sim$20\% in Scenario F.

\subsection{Implications for the Habitable Worlds Observatory} \label{subsec:HWOGoalsImpact}

To minimize the yield reduction from meteorite impact and system degradation over time for HWO, several possible mitigation measures could be considered. This may involve adjustments to key system components, such as the barrel design, aperture size, or the selection of detectors and coronagraphs. At the same time, a certain level of yield loss should be anticipated and accounted for in both mission planning and performance margins. Considering the reduction in yield expected from micrometeoroid damage in this study, we present several possible mitigation strategies in this section. 

A significant reduction in yield loss could be achieved by using mitigations that lower the probability that the mirror is impacted by high energy particles. These mechanisms include a longer barrel design, which physically shields the mirror, and the implementation of on-orbit pointing constraints that reduce high velocity impactors. The largest energy single-hit impact event presented in this study is the 5 year RAM pointing impactor from Table \ref{tab:MMImpactMasses} using the shortest barrel design (77.78 J), the MR model for crater diameter, and the Scenario A telescope design \citep[from ][]{Stark2024yield}, which results in a $\sim$60\% reduction in yield. \revision{Swapping in the longest barrel design impactor case in the 5 year timeline (45.74 J) reduces this to about 50\% reduction in yield, illustrating how selecting longer barrel designs can improve telescope performance over time} 
\revision{Furthermore, none of the Anti-RAM only pointing impactor cases result in a significant reduction to yield, thus instituting a micrometeoroid avoidance zone (MAZ) that pushes the maximum probable impactor towards the teal region of Figure \ref{fig:YieldCurves} will greatly help to protect the primary mirror.}


Yield reduction can be mitigated further by improving the throughput of the telescope and instrument which makes the system more robust to damage-induced losses. For example, if we conduct the same exercise in the previous paragraph using the Scenario F design which, among many improvements, boasts a coronagraph with a higher throughput than Scenario A \citep{Stark2024yield}, the worst case reduction in yield would be a \revision{$\sim$55\%} loss (compare to 60\% in Scenario A), which could be lowered to \revision{$\sim$45\%} with the longest barrel design. With a baseline expected yield of 43.2, Scenario F already offers a higher anticipated yield than Scenario A (baseline of 16.6), making yield reductions more tolerable. Not only does Scenario F offer a higher baseline yield, but the higher throughput coronagraph is less sensitive to stray light overall due to the increased planet signal, meaning the anticipated reductions in expected yield are lower for the same impact event with Scenario A (compare 55\% in Scenario F to 60\% in Scenario A for the worst case). \revision{Finally note, the baseline yield for a 2 year survey duration in Scenario A (16.6) is already below the required science yield of the mission (25), so if this design case was selected the survey duration would need to be increased, effectively raising the probability of problematic impact events.}

In the previous section, when inferring damage from a 26 J impactor (the higher end of the range found for the C3 JWST event), we found an expected yield reduction of 20\% for Scenario F.  \revision{Further study is required to determine if the Heliospace test is appropriate to calibrate yield loss estimates to; if that is the case,} and the BRDF used in this study is a reasonable approximation, our work suggests we may need to cautiously budget for at least an expected 20\% loss in yield due to micrometeoroid damage. However, we caution that BRDF measurements require optical scatter measurement and analysis, which will necessitate many impact tests. Thus, improving confidence in the estimated yield loss we should budget for will require further extensive impact testing. \revision{Note, the estimates in this analysis are likely optimistic, as the tests do not include a coated mirror sample or account for performance degradation in the coating proximate to the damage.}

More innovative, and perhaps longer-term,  approaches could involve developing mitigation protocols for HWO  that reduce stray light following a high energy impact event. This could potentially center around making the damage crater less reflective, or removing or replacing damaged mirror segments. Having noted this as a possible avenue, we leave the study of such a protocol to future efforts.

Overall, to ensure the science goals of HWO are met, \revision{we strongly suggest that the imapact to the science yield from damage be considered during the exploration of the various architectural options.  Additionally, }we suggest establishing a MAZ from the start of the mission, adopting the longest feasible barrel design to prevent high energy low probability impacts, and pursuing a high-throughput system design that has sufficient margins that the science goals can still be achieved even with moderate meteoroid damage. Future work should focus on the largest source of uncertainty: the relationship between crater diameter and impact energy. \revision{Moreover, these studies should include all aspects of the HWO flight mirror design, materials, chamfers, coatings and surface finish. }

\subsection{Limitations of the Study}\label{subsec:StudyLimitations}

In this work we quantified the order of magnitude of stray light and approximate reduction in yield that may be expected for single-hit, high energy micrometeoroid impacts. There are many considerations that we have not explored that could affect the stray light expected from micrometeoroid damage for HWO \revision{and are summarized below.}

\revision{In this work, we assumed all meteoroids had a density of 2.5 g/cm$^{3}$ and an impact velocity of 20 km/s. It may be informative to change these constraints to ranges of densities and velocities. However, for a given impactor density, the Watts and MR damage models used here predict a crater diameter that is ultimately a function of impact energy. Thus, mass and impact velocity pairings that yield the same impact energy will predict the same yield loss for a fixed meteoroid density. Assuming faster impact velocities will mean less massive particles will be more problematic than found by the current study. This work was meant to provide a standardized analysis to determine if micrometeoroid damage may constitute a significant stray light source, not be a fully comprehensive exploration of possible impact events.}

We focused exclusively on worst-case single-hit impacts and did not address the cumulative effect of damage building up from many smaller energy events. For JWST, minor micrometeoroid strikes occur with a frequency of 1 -- 2$\times$ per month, consistent with pre-launch expectations \citep{Moorhead2020replicate,Rigby2022characterization,Menzel2023jwstreport,MenzelC3Energy}. When considering the Grun model for the sporadic micrometeoroid environment, it was found that impacts with energy $\geq$0.5 J may be expected on a cadence of $\sim$2$\times$ per month \citep{grunModel,MenzelC3Energy}. Considering the 8 J Heliospace impact test introduced previously (\S\ref{subsec:ImpactTestDiscuss}) led to relatively tolerable reductions in yield of 2\% -- 4\% in the single-hit impact case, our work suggests that single-hit impacts of 0.5 J would have a negligible effect on science yield (Figure \ref{fig:YieldCurves}). However, considering these minor impacts are occurring on a monthly basis, the cumulative build up of damage may affect science yield. A future study focused on this question could be conducted with the tools developed in this work. 

In terms of modeling strategy, this study adopts an approach that maximizes computational efficiency by using a semi-analytic approach to calculate stray light, though this comes at the expense of the greater sophistication offered by more detailed ray-tracing methods. To implement the yield calculator in our work, it was necessary to calculate stray light for a large number ($\sim$1000) of sources, over an array of impact energies, and using two separate models relating crater diameter to impact case. This led to over 5$\times$10$^{4}$ calculations of stray light observations, motivating our computationally efficient approach. Through approximating stray light, we have demonstrated that this particular noise source should not be ignored. The next step is then to dedicate resources to provide more accurate estimates of the expected damage from various impactors on materials relevant to HWO, and to conduct more refined ray tracing calculations for the most problematic impact events identified. \revision{Moreover, the reflectivity of impact craters may change significantly, and future work in both impact testing and modeling should test the expected reduction in reflectivity.} In contrast to analytical BRDF models, ray-tracing methods provide better accuracy by accounting for complex geometries and optical paths created by realistic and imperfect baffle and barrel structures. Additionally, ray-tracing models can account for multiple bounces of light and more complicated surface compositions.

While we have provided some anchoring for expected yield reductions from the estimated highest energy impact events anticipated for HWO, further consideration is needed to define the maximum impact energy the mission should be designed to withstand. Considering the maximum 2$\sigma$ probability impactors presented in \S\ref{subsec:MeteorDamModel}, only 2 directions were considered, RAM and Anti-RAM, with highest and lowest impactor velocity distributions, respectively. During telescope operation, RAM direction pointing will likely be minimized using a meteor avoidance zone constraint, and the telescope will also rarely be pointed perfectly in the most favorable Anti-RAM direction. Thus, the true maximum energy impactors will fall between these pointing cases for HWO. 
\revision{It is also worth noting that the Altruistic Yield Optimizer (AYO) \cite{Stark2014AYO} used in this study does not currently account for any pointing restrictions. If the yield calculator included a MAZ, it would complicate the scheduling of revisits and may cause an overall reduction in yield. This addition will further require significant development to AYO, which we leave to a future study.}

Finally, we note that micrometeoroid damage will not be the only contributor of stray light. In our yield results we have allowed the entire stray light budget of the mission to be attributed to micrometeoroid damage, and stray light from a binary companion due to approximate mirror roughness and contamination \citep[as implemented in][]{Stark2024yield}. This is neglecting several other sources of stray light that may be significant, including reflections off of different parts of the observatory (e.g., chamfer, barrel sides). This means that our results are optimistic -- allowing for higher levels of micrometeoroid damage for a given reduction in yield \revision{than are likely to apply for the HWO mission.}

\section{Conclusions} \label{sec:Conclusions}

This work has explored order of magnitude estimates of the stray light and resulting reduction in exoEarth candidate yield expected for various single-hit micrometeoroid impact events for an HWO-like mission. As a complement to state-of-the-art ray tracing models for calculating stray light, we have developed a computationally efficient model that relies on analytical techniques to consider a vast sample of targets and impactor energies. Even for targets near the galactic plane, we have shown that the stray light from the host star will dominate the cumulative stray light contributed by micrometoroid impact events for impact energies less than 2 mJ; for impact events less energetic than this, considering the full sky background around the target of interest may increase the stray light resulting from the impact by up to 3 -- 4 orders of magnitude, at most. 

The assumed relationship between the impact energy and size of the resulting damage crater has been shown to have a large effect on the total stray light. The two bounding case models we explored in this work (Watts and MR) led to stray light predictions that differed by several orders of magnitude. This further translates to a broad difference in the expected reduction on yield. The Watts model, which generally predicts smaller craters for the same size impact than the MR model, would predict no effect on yield for impact energies $<$200 J, while the MR model would predict no effect on yield for impact energies $<$0.2 J. An initial study of an 8 J impact by Heliospace (\S\ref{subsec:ImpactTestDiscuss}) suggested the true relationship between damage diameter and impact energy will fall between the Watts and MR models. 

To increase the probability that HWO will achieve its exoEarth candidate yield of 25 \citep{national2021decadal}, we recommend adopting the longest feasible barrel design -- to reduce the risk of high energy, low probability impacts -- and implementing a micrometeoroid avoidance zone (MAZ) from the start of the mission. Informed by the 8 J impact test by Heliospace, our work suggests an impact similar to the May 2022 event on JWST's C3 mirror segment \citep[7 -- 26 J][]{Menzel2023jwstreport,MenzelC3Energy} may produce a reduction in exoEarth candidate yield of 2\% -- 30\%, depending on the exact impact energy and coronagraph design. If a high-throughput coronagraph is adopted, we recommend that HWO budget for an expected 20\% reduction in yield due to micrometeoroid damage. Alternatively, loss in yield may be prevented by creating some damage mitigation protocol to be applied after a high energy impact event. 

Overall, our work shows that the potential stray light contributed by the highest energy micrometeoroid impact events expected for HWO cannot be ignored at this stage in the mission design process, and may have large effects (\revision{up to 60\% reduction}) on the science yield of the mission. Future studies are required to provide more accurate estimates of maximum possible impact energies expected during the mission, the contribution of stray light from the resulting damage craters, and especially the relationship between crater size and impact energy for materials relevant to HWO. 

\section*{Disclosures}

The authors declare there are no financial interests, commercial affiliations, or other potential conflicts of interest that have influenced the objectivity of this research or the writing of this paper.

\section*{Code, Data, Availability}
All code used to complete this work can be found on GitHub in the following repository:

https://github.com/mgialluca/StrayLightBackgroundModel

\section* {Acknowledgments}

We would like to thank Jeffery Puschell and Tiffany Glassman \revision{of Northrop Grumman} for providing advice and support, which greatly helped to make this collaboration possible. We further thank Joe Pitman and Heliospace for sharing the results of their impact test, this information greatly improved the scientific value of our study. We would also like to thank Len Seals for proof reading the final manuscript. \revision{Finally, we thank 3 anonymous peer reviewers for their suggestions which greatly improved the quality and clarity of our work.}

This work was completed during a University of Washington Astrobiology Program research rotation; M.G. completed work at both Northrop Grumman Corp and NASA Goddard Space Flight Center in an equal time split.
M.G. acknowledges funding from the NSF Graduate Research Fellowship (DGE-2140004), as well as the University of Washington's Astrobiology Program. This work was supported in part by the Virtual Planetary Laboratory Team, a member of the NASA Nexus for Exoplanet System Science, and funded via NASA Astrobiology Program Grant 80NSSC18K0829. This work made use of the advanced computational, storage, and networking infrastructure provided by the Hyak supercomputer
system, which was funded by the University of Washington and the Virtual Planetary Laboratory.

TDR gratefully acknowledges support from the Cottrell Scholar Program administered by the Research Corporation for Science Advancement and from NASA's Habitable Worlds Program (No.~80NSSC20K0226). \revision{JWA and BS were supported on Northrop Grumman internal funding.}


\bibliography{report}   
\bibliographystyle{spiejour}   




\appendix

\section{Stray Light Model Description} \label{sec:FullSLModelDescriptionAppendix}

\begin{figure}
    \centering
    \includegraphics[width=\textwidth]{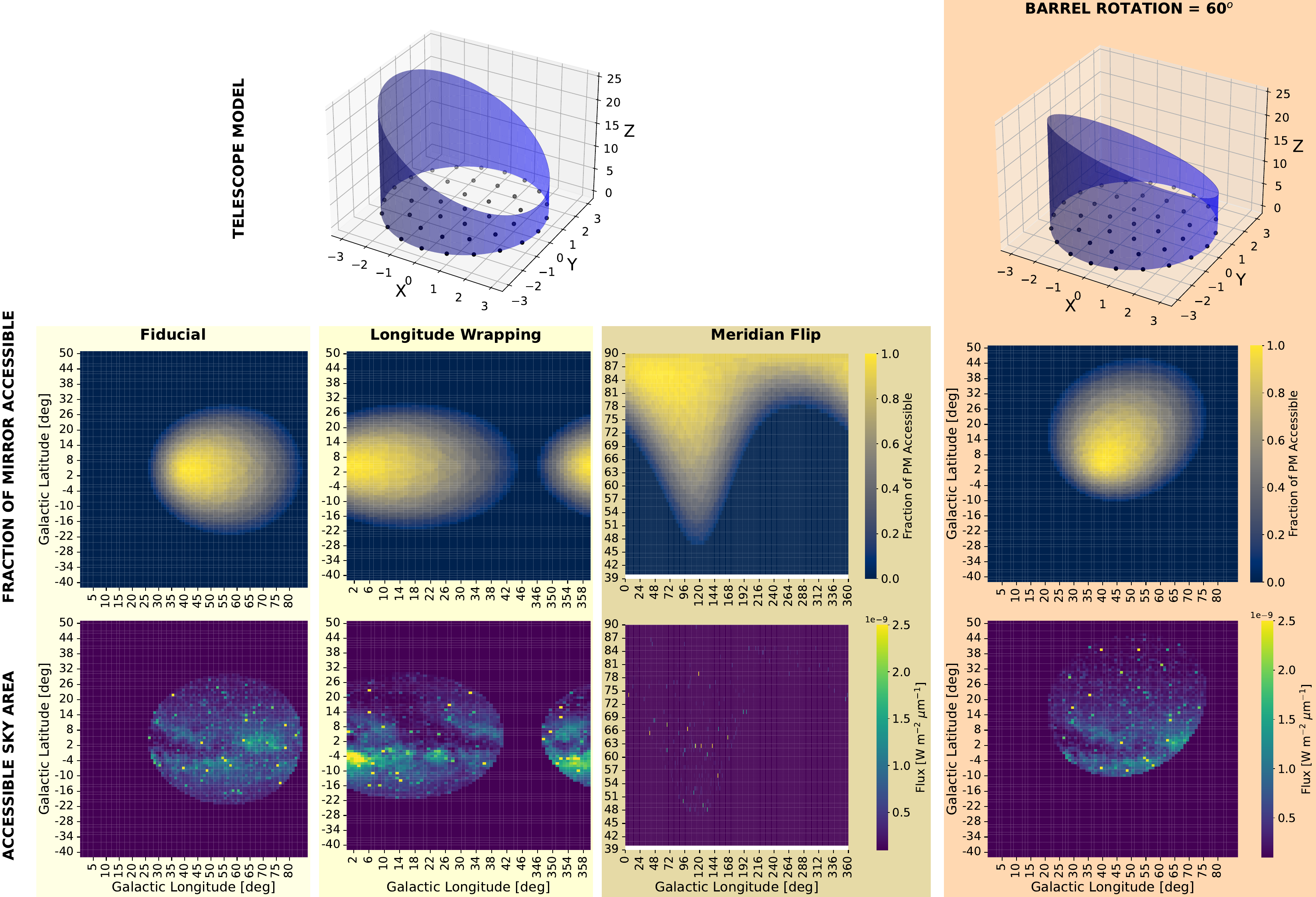}
    \caption{Several examples of the stray light model for a 6 m telescope with a scarfed barrel whose length goes from 4x the diameter on one side, to 1x the diameter on the other. The top row shows the 3D telescope model for a 0$^{o}$ [left] and 60$^{o}$ [right] barrel rotation. The middle row shows the fraction of the mirror accessible to each sky point, or the mirror coverage fraction (MCF). The bottom row shows the accessible sky area, or the sky points accessible to each observational setup. The first 3 columns apply the telescope model with a 0$^{o}$ barrel rotation. The first column shows the fiducial setup, which uses a pointing direction of (40$^{o}$, 5$^{o}$) in galactic longitude and latitude, respectively. The second column shows the ``Longitude Wrapping" case, which uses a pointing direction of (0$^{o}$, 5$^{o}$) to demonstrate the model's ability to handle wrapping in galactic longitude from 0$^{o}$ to 360$^{o}$. The third column shows the ``Meridian Flip" case, which uses a pointing direction of (40$^{o}$, 85$^{o}$) to demonstrate the model's ability to handle the meridian flip that occurs at pointing directions near and at the galactic poles. Finally, the last column [light orange background] illustrates how the fiducial case would change for a barrel rotation of 60$^{o}$.}
    \label{fig:SLModelExample}
\end{figure}

Here we describe the process implemented in our stray light model, which can be accessed on GitHub\footnote{https://github.com/mgialluca/StrayLightBackgroundModel}. To preserve computational efficiency during a stray light estimation, the model first reduces the total number of sky points by truncating the sky map to the approximate area visible to the primary mirror. To do so, we extract a circular field centered on the pointing direction with radius equal to the maximum angular separation any point on the primary mirror may be able to access before its line of sight is blocked by the telescope barrel. This maximum angular separation is determined by the edge of the primary mirror directly across from the shortest side of the telescope barrel and can be described by:
\begin{equation}
    \tan \theta_{i, {\rm max}} =  \frac{D_{\rm PM}}{l_{\rm short}}  \ ,
\end{equation}
where $l_{\rm short}$ is the length of the shortest side of the barrel. Once a truncated patch of the sky background has been extracted, the model will iterate through every point in the patch, determine which mirror surface points have an optical path to that sky point (that is, not obstructed by the barrel), and calculate the BRDF model contribution for each sky-mirror point pair (e.g., evaluating the integral of Equation \ref{eq:FullSLRequire}). Figure \ref{fig:SLModelExample} illustrates several steps of this process, including the 3D telescope model (top row), the fraction of mirror each sky point has access to (i.e., the "mirror coverage fraction", or MCF; middle row), and the sky map patch relevant for that observation.

Due to the predominantly specular nature of the BRDF models used to describe defective mirror area, the closer a sky point is to the pointing direction, the larger its contribution to the cumulative stray light. Because of this, the the sky map must be super-resolved near the pointing direction. Recall, each point on the sky map is the cumulative flux from all \textit{Gaia} sources that fall in that point's grid box on-sky; the assumption being that the total contribution from each source in a sky grid box is approximately the same as summing the flux from those sources and calculating a single contribution using this summed flux and the incidence direction defined by the center of the box. For the majority of the sky map, this is true. However, sources in the grid box containing the pointing direction cannot all be approximated as possessing a 0$^{o}$ incidence angle, lest the cumulative stray light contribution be significantly overestimated. For the sky grid box that contains the pointing direction, the contribution from each source in the box is individually resolved and calculated; requiring an additional Gaia query to return a list of the resolved sources in the pointing box including their coordinates and fluxes in the bandpass of interest.

\subsection{Reference Frame Transformations} \label{subsec:refframetransforms}

To fully describe reference frame transformations that are used in our model, we first outline the default description of how the optical axis reference plane relates to the pointing direction and sky coordinates. Picture the xyz-plane of the telescope model (shown by the 3D plots in the top row of Figure \ref{fig:SLModelExample}) projected onto the sky plane; the z-axis vector is the pointing direction, the x-axis is positive for increasing galactic longitude, and the y-axis is positive for increasing galactic latitude. Observation targets are considered to be sufficiently far away that the incidence direction ($\theta_i$, $\phi_i$) is the same for any point on the mirror with respect to the direction given by the normal of the mirror's center point (which lies along the z-axis). 

For the evaluation of an individual sky-mirror point contribution, the integration over the solid angle of reflection becomes a double integral over the spherical coordinates $\theta_r$ (angle from the surface normal to the direction of the secondary mirror) and $\phi_r$ (azimuthal angle from the mirror plane's x-axis). The bounds of integration can be unclear for non-zero reflectance angles, but when $\theta_{r}$ = 0$^{o}$ the bounds of integration over the planet's PSF core are simply 0 to $\lambda/D$ for $\theta$ and 0 to 2$\pi$ for $\phi$. To maintain these bounds of integration, we switch reference frames when evaluating the integral to that in which the vector pointing towards the secondary mirror (i.e., the reflectance direction, $\Vec{V_{r}}$) lies along the z-axis. In other words, $\theta_{r}$ is transformed to $\theta_{r}'$, and the relationship between the reflectance direction and every other vector in the scene is preserved; the other vectors of importance primarily include the direction of incidence ($\Vec{V_{i}}$), the mirror area point dA's normal vector ($\Vec{n_{dA}}$), and the direction of perfect specular reflection ($\Vec{V_{ps}}$, defined by $\Vec{V_{i}}$ and $\Vec{n_{dA}}$). 

\begin{figure}
    \centering
    \includegraphics[width=\textwidth]{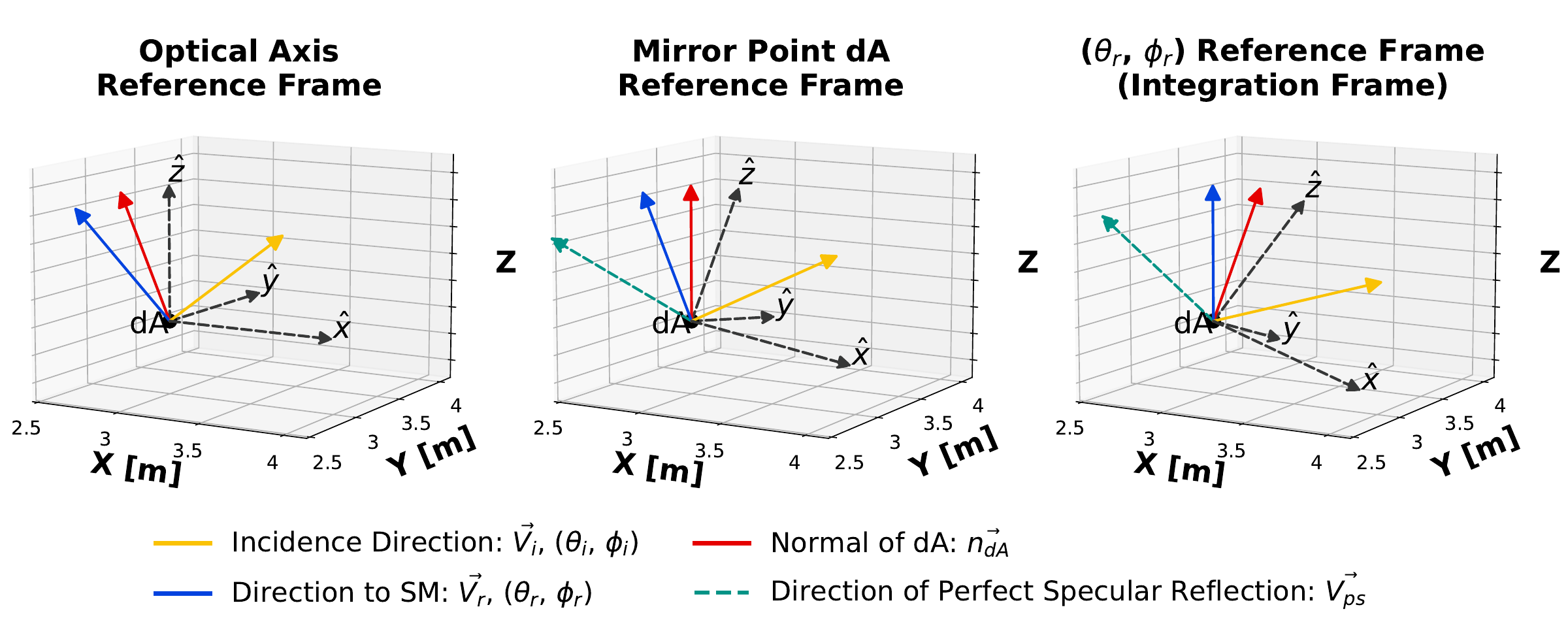}
    \caption{An example of the reference frame transformations used in our stray light model for a mirror point, dA, located at (x,y) position (3 meters, 3 meters) from the mirror center point located at (0, 0), with incident light from the direction ($\theta_{i}$, $\phi_{i}$) of (45$^{o}$, 0$^{o}$) and the secondary mirror located 7 meters above the primary (similar to JWST). The length of arrows here are arbitrary, and mainly used to show direction (hence the Z axis is unit-less). In all subplots, the red arrow indicates the mirror point dA's normal vector ($\Vec{n_{dA}}$), the blue arrow indicates the reflection direction to the secondary mirror ($\Vec{V_{r}}$), and the yellow arrow indicates the direction of incidence ($\Vec{V_{i}}$). [Left]: Shows the original, or optical axis, reference frame defined by the mirror center point; across all three subplots, $\hat{x}$, $\hat{y}$, and $\hat{z}$ are the xyz unit vectors from the original reference frame. [Middle]: The reference frame of the mirror point dA, such that $\Vec{n_{dA}}$ is on the z-axis; in this frame, the direction of perfect specular reflection (teal dashed line, $\Vec{V_{ps}}$) is defined based on the orientation of $\Vec{V_{i}}$ to $\Vec{n_{dA}}$. [Right]: The reference frame of $\Vec{V_{r}}$ (such that it would line up with the z-axis); this is the reference frame in which integration is performed to simplify the bounds of integration for $\theta_{r}$ and $\phi_{r}$.}
    \label{fig:refframetransform}
\end{figure}

Figure \ref{fig:refframetransform} shows an example for how these reference frame transformations are conducted in the model. In this example, the mirror point is located at (x,y) = (3 meters, 3 meters) from the mirror center point at (0,0), the incidence direction ($\theta_i$, $\phi_i$) is (45$^{o}$, 0$^{o}$), and the secondary mirror is located 7 meters above the primary (similar to JWST). The right panel of Figure \ref{fig:refframetransform} shows the reference frame in which the integration would be performed (where $\Vec{V_{r}}$ is aligned with the z-axis).

When evaluating Equation \ref{eq:FullSLRequire} in a transformed reference frame, several directional variables need to be modified to compute the stray light contribution correctly. First, for a particular reflectance direction, the BRDF is multiplied by a factor of $cos(\theta_{r,dA})$ before integration over the solid angle, as can be seen in the derivations of \S\ref{sec:analyticApproach}. To preserve the value of this factor in any integration reference frame, we redefine this as the dot product of the reflectance direction, $\Vec{V_{r}}$, and the vector normal to dA, $\Vec{n_{dA}}$:
\begin{equation} \label{eq:costhetar}
    cos(\theta_{r,dA}) \rightarrow \Vec{V_{r}} \cdot \Vec{n_{dA}} \ ,
\end{equation}
which is by definition the angle of reflectance, $\theta_{r,dA}$. So long as $\Vec{V_{r}}$ and $\Vec{n_{dA}}$ are defined in the same frame of reference, Equation \ref{eq:costhetar} will always yield a consistent result. In the frame of integration used here, $\Vec{V_{r}}' \cdot \Vec{n_{dA}}'$ is used in the integration where $\Vec{V_{r}}'$ is centered at $\theta_{r}'$ = 0$^{o}$ as shown in the right panel of Figure \ref{fig:refframetransform}.
 

\revision{The Peterson BRDF model used in this work \cite{Peterson2012brdf} is dependent on the direction of perfct specular reflection (for a given incidence direction). This dependence appears as the difference of the directional vector of reflection and that of perfect specular reflection ($|\Vec{V_{r}}$ - $\Vec{V_{ps}}|$).}
Perfect specular reflection occurs when $\theta_{i,dA}$ = $\theta_{r,dA}$, and the difference in azimuthal angle is 180$^{o}$ ($|\phi_{r,dA}$ - $\phi_{i,dA}|$ = 180$^{o}$); this fact needs to be preserved in any frame of reference to ensure specular reflection is properly treated.

To preserve the relationship between the reflectance direction and that of perfect specular reflection in the reference frame used for integration, 
our model first defines the direction of perfect specular reflection, $\Vec{V_{ps}}$, by mirroring $\Vec{V_{i}}$ (i.e., $\theta_{ps}$ = $\theta_{i,dA}$ and $\phi_{ps}$ = $\phi_{i,dA}$ + 180$^{o}$) in the mirror point dA's reference frame, the frame where the normal of dA is pointed along the z-axis. This is illustrated in the middle panel of Figure \ref{fig:refframetransform}. Then $\Vec{V_{ps}}$ transforms to $\Vec{V_{ps}}'$ in the integration reference frame. The dot product of $\Vec{V_{ps}}'$ with $\Vec{V_{r}}'$ (centered about $\theta_r'$ = 0$^{o}$) gives the angle between the reflectance direction and that of perfect specular reflection in the integration, and the difference between $|\Vec{V_{r}}'$ - $\Vec{V_{ps}}'|$ will be equivalent in the integration reference frame to the difference of the two directions in any other reference frame.

\listoffigures
\listoftables

\end{spacing}
\end{document}